\pdfoutput=1
 \documentclass[fleqn,final,5p,times,twocolumn]{elsarticle}
\usepackage[T1]{fontenc}		% font encoding
\usepackage[utf8]{inputenc}		% input encoding
\usepackage{lmodern}			% fonts
\usepackage[english]{babel}
\usepackage{booktabs}	% provides nice tables

\usepackage[usenames,dvipsnames]{xcolor} % provides colours

\usepackage[colorlinks]{hyperref}

\usepackage{graphicx}
\usepackage{tikz,pgfplots}		% provides TikZ
\usetikzlibrary{arrows,shapes}

\newlength\figureheight
\newlength\figurewidth

\usepackage{subfig}				% provides subfloat

\usepackage{amsmath}

\usepackage{amssymb}
\usepackage[squaren]{SIunits}	% provides \metre, etc.

\usepackage{array}

\newcommand*{\mycdot}{\kern-.2em\cdot\kern-.2em}
\newcommand{\rvec}{\textbf{r}}

\newcommand{\uvec}{\textbf{u}}

\newcommand\copyrighttext{%
  \scriptsize \textcopyright\, 2016. This manuscript version is made available under the CC-BY-NC-ND 4.0 license
\url{http://creativecommons.org/licenses/by-nc-nd/4.0/}.}
\newcommand\copyrightnotice{%
\begin{tikzpicture}[remember picture,overlay]
\node[anchor=south,yshift=10pt] at (current page.south) {\fbox{\parbox{\dimexpr\textwidth-\fboxsep-\fboxrule\relax}{\copyrighttext}}};
\end{tikzpicture}%
}

\usepackage[normalem]{ulem}

     \newcommand{\myHighlight}[1] {#1}
     \newcommand{\myDelete}[1] {}

\journal{Chemical Engineering Science}
\begin{document}
\begin{frontmatter}
\title{Recurrence CFD -- a novel approach to simulate multiphase flows with strongly separated time scales}

\author[rvt]{T. Lichtenegger\corref{cor1}}
\ead{thomas.lichtenegger@jku.at}

\author[rvt]{S. Pirker}
\ead{stefan.pirker@jku.at}

\cortext[cor1]{Corresponding author}

\address[rvt]{Department for Particulate Flow Modelling, Johannes Kepler University,
Linz, Austria}

\begin{abstract}
Classical Computational Fluid Dynamics (CFD) of long-time processes with strongly separated time scales is computationally extremely demanding if not impossible. Consequently, the state-of-the-art description of such systems is not capable of real-time simulations or online process monitoring.

In order to bridge this gap, we propose a new method suitable to decouple slow from fast degrees of freedom in many cases. Based on the recurrence statistics of unsteady flow fields, we deduce a recurrence process which enables the generic representation of pseudo-periodic motion at high spatial and temporal resolution. 

Based on these fields, passive scalars can be traced by recurrence CFD. While a first, Eulerian Model A solves a passive transport equation in a classical implicit finite-volume environment, a second, Lagrangian Model B propagates fluid particles obeying a stochastic differential equation explicitly.

Finally, this new concept is tested by two multiphase processes – a lab scale oscillating bubble column and an industrial scale steelmaking converter. Results of tracer distribution obtained by recurrence CFD are in very good agreement with full CFD simulations, while computational times are dramatically reduced. Actually, recurrence CFD is a promising candidate for online simulations of passive transport processes at full CFD resolution, which opens the door towards improved process monitoring. 
\end{abstract}
\end{frontmatter}
\copyrightnotice
% ----------------------------------------------------------------------
\section{Introduction}
\label{S:Intro}
Computational Fluid Dynamics (CFD) can be regarded as an established investigation and design tool ranging from academic to industrial purposes. Dedicated multiphase models enable the simulation of complex, industrial-scale flows, e.g.\ polymerisation in fluidized beds or steel refining in gas stirred vessels. Due to the highly dynamic nature of these multi-scale processes, CFD simulations typically require small time-steps as well as \myDelete{large}\myHighlight{great} spatial resolution. Consequently, they are nearly always associated with large computational times, which limits their use to short-term, offline investigations.

While numerous flavors of coarse graining (e.g.\ \cite{Heynderickx2004,Igci2008,Igci2011b,Igci2011,Milioli2013,Ozel2013,Parmentier2012,Schneiderbauer2013,Schneiderbauer2014,Wang2008,Wang2010,Yang2003,Zhang2002}) with different levels of sophistication have advanced multiphase CFD simulations to industrial scales, their utilization for the description or even online monitoring of transient, long-term processes requires further, dramatic simplification of the solution procedure.

Various strategies aiming at decreasing the complexity of either the spatial or temporal representation or both have been proposed in the literature.
Temporal aspects of highly dynamic CFD simulations may be simplified by replacing unsteady flows with their time-averages for the purpose of tracing passive scalars, e.g.\ concentrations, on steady fields. However, care must be taken because averaging the volume fraction fields of multiphase flows can lead to completely unphysical results. If, for instance, the unsteady propagation of a bubble in a fluidized bed is not resolved, gas-solid contact is not directly accessible.

In less heterogeneous systems, on the other hand, consequences are expected to be less severe so that time-averaging has been employed as a basis for further simplifications. For example, \cite{Rigopoulos2001,Rigopoulos2003} simulated bubble column reactors by dividing the computational domain into a series of large compartments which they assumed to be perfectly stirred with homogeneous properties. Conventional CFD simulations were then used to determine the mass exchange rates between those compartments as well as the mean rates of change within each of them.
While their compartmentation approach results in a very efficient representation of flow processes, it relies on the validity of time-averaging and it inevitably loses structural information.

Less brutal reduced order models (ROM) try to describe the essential dynamics of complicated flows in a low-dimensional fashion but go beyond mere time-averaging. The full problem's huge amount of degrees-of-freedom is drastically reduced by representing flow fields as a superposition of their most energetic modes obtained from Proper Orthogonal Decomposition (POD). The time-evolution of each modes' weight can then be calculated with a truncated Galerkin projection approach (\cite{Aubry1988,Berkooz1993,Lumley1996,Sirovich1987}).
This procedure has proven to be a powerful simulation tool for various fluid mechanical problems (\cite{Akhtar2009,Bergmann2005,Cazemier1998,Deane1991,Liakopoulos1997,Rempfer1994,Siegel2008,Singh2001,Wang2012b}) including multiphase flows (\cite{Brenner2010,Brenner2012,Cizmas2003,Yuan2005}). However, it is well-known that long-time integration of POD-modes may predict spurious asymptotic behavior (\cite{Foias1991,Rempfer2000,Sirisup2004}). Besides other methods involving several empirical parameters to control stability (an overview and further references are given by \cite{Bergmann2009}), such a failure can be avoided at the cost of repeatedly supplementing the POD-governed time evolution of the mode coefficients with time intervals of full-model integration (\cite{Bergmann2009,Rapun2010,Sirisup2005}), e.g.\ the Navier-Stokes equations. This leads to a remarkable speed-up in comparison to ordinary simulations, but the method is still limited by repeated, time-consuming full CFD calculations.
Particularly for situations where long-time properties are of central interest, these may be a bottleneck.

Since such systems, if not converging to a steady state, are often characterized by recurring patterns, we propose a different, more economic approach based on the recurrence analysis introduced by \cite{Eckmann1987}.
While running a full CFD simulation, we store the entire flow field at a monitoring frequency $f_{\text{rec}}$ for a monitoring period $\tau_{\text{rec}}$. Next, a recurrence statistics (see next section) is constructed by comparing fields at two times with each other based on a given norm. With this information it is possible to judge to which extent states at different times are similar. Consequently, a generic, unsteady flow can be composed by threading individual flow sequences stored beforehand.

It turns out that due to this procedure, the method suffers from local (but not global) violation of conservation laws whenever a new sequence is artificially stitched to another one. We discuss this issue by presenting the solution of a conventional transport equation (Model A in the following section). Alternatively, we propose Model B which overcomes this violation of conservation by relaxation towards the recurrence fields in a conservative way. Finally, this flexibility allows for the combination of different process states which are represented by individual recurrence statistics.

In the following section we introduce our concept of recurrence statistics and the generation of generic flow fields. Next, the two recurrence CFD models are presented. They are thoroughly discussed by applying them to an oscillating bubble column. Finally, Model B is applied to an industrial process of steelmaking, where we also consider changing process conditions. In the outlook, existing limitations and future potential of recurrence CFD are discussed.

% ----------------------------------------------------------------------
\section{Modelling}
\subsection{Recurrence Statistics}
A recurrence statistics contains information about a system's degree of similarity at two different times over a certain time range $\tau_{\text{rec}}$ resolved with a frequency $f_{\text{rec}}$. For some cases, these parameters can be estimated a priori. In general, however, simulation data need to be analyzed to obtain sensible values. As an example, we show in section~\ref{sec:obc}, Figs.~\ref{fig:sigAlpha} -- \ref{fig:sigVz} the phase fraction and the velocity of a liquid stirred by an oscillating bubble column probed at a specific point. Though clearly visible to the eye in this case, spectral analysis of the signals as displayed in Fig.~\ref{fig:sigFourier} would reveal their (pseudo)periodicity also in less obvious situations. To obtain a useful recurrence statistics, its time range needs to span several pseudo-periods $\tau_{\text{p-p}}$ corresponding to the lowest-lying peak frequency $f_{\text{p-p}}$, i.e.
\begin{equation}
 \tau_{\text{rec}} \gg \tau_{\text{p-p}} = \frac{1}{f_{\text{p-p}}}.
\end{equation}
Although, on the contrary, the highest frequency necessary to resolve could also be determined from Fourier analysis, we estimate it from a more practical point of view. Given a signal $\varphi(t)$, the relative error in its interpolated values between two sampling times $t$ and $t+\Delta t$ is small if
\begin{equation}
  |\varphi(t)| \gg |\varphi(t+\Delta t)-\varphi(t)| \approx \Delta t |\dot{\varphi}(t)|.
\end{equation}
For a general $\varphi(t)$ containing zeros, the consequential criterion
\begin{equation}
 \Delta t \ll \frac{|\varphi(t)|}{|\dot{\varphi}(t)|}
 \label{eq:tref0}
\end{equation}
cannot be fulfilled for all times $t$. However, the choice
\begin{equation}
 \frac{1}{f_{\text{rec}}}\equiv \Delta t_{\text{rec}} \ll \sqrt{\frac{\langle \varphi^2 \rangle}{\langle \dot{\varphi}^2 \rangle}}\equiv \frac{1}{f^{(\varphi)}_{\text{crit}}}
 \label{eq:tref}
\end{equation}
with $\langle \cdot \rangle$ denoting time-averaging guarantees that it holds most of the time. Note that if the spectrum of $\varphi(t)$ is limited by $f_{\text{max}}$, it is easy to show that $f_{\text{max}}$ satisfies Eq.~\eqref{eq:tref}.
With more than a single signal available, the sampling frequency has to be chosen such that it is larger than the largest critical frequency from all signals.

\begin{figure}[t]
\centering
\subfloat[\label{fig:recMat_alpha}]{%
  \includegraphics[height=0.3\textwidth]{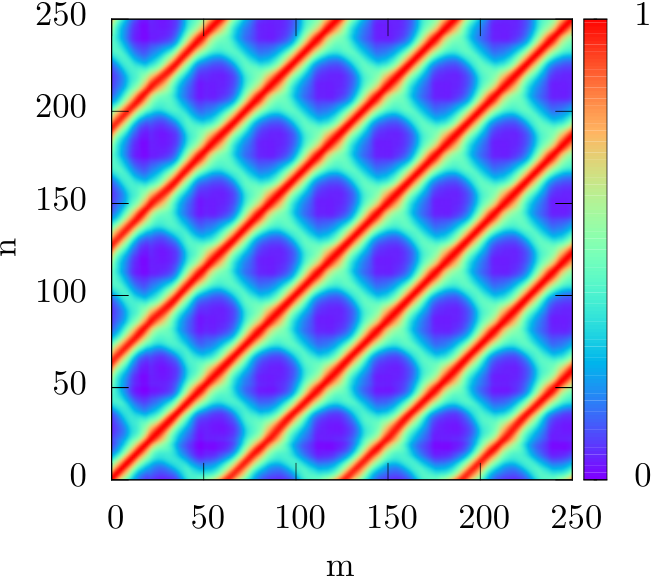}
}\\
\subfloat[\label{fig:recMat_phi}]{%
\includegraphics[height=0.3\textwidth]{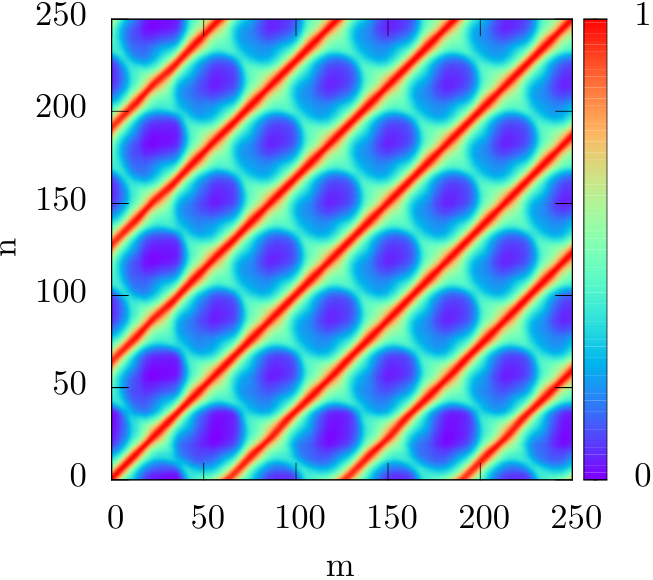}
}
\caption{Recurrence plots for matrices (a) $R^{(\alpha)}_{m,n}$ and (b) $R^{(\phi)}_{m,n}$ of 250 snapshots of an oscillating bubble column. Anti-diagonal local maxima corresponding to motion in opposite directions are slightly more pronounced for $R^{(\alpha)}_{m,n}$ than $R^{(\phi)}_{m,n}$.}
\end{figure}

Returning to the case of a multiphase flow with volume fractions $\alpha^{(i)}(\rvec,t)$, velocity fields $\uvec^{(i)}(\rvec,t)$ and fluxes ${\boldsymbol{\phi}^{(i)}(\rvec,t)\equiv \alpha^{(i)}(\rvec,t)\uvec^{(i)}(\rvec,t)}$, a norm 
to quantify the similarity of states at $t$ and $t'$ is needed. There is by no means a unique or ``best'' expression for all cases.  If there is one dominant phase, e.g.\ $i=1$, we intuitively suggest to define functions
\begin{align}
 R^{(\alpha)}(t,t')&\equiv 1 - \frac{1}{{\cal N}^{(\alpha)}}\int d^3r \big(\alpha^{(1)}(\rvec,t) - \alpha^{(1)}(\rvec,t')\big)^2 \label{eq:recmatdef1} \\
  R^{(\phi)}(t,t')&\equiv 1 - \frac{1}{{\cal N}^{(\phi)}}\int d^3r \big(\boldsymbol{\phi}^{(1)}(\rvec,t) - \boldsymbol{\phi}^{(1)}(\rvec,t')\big)^2,
 \label{eq:recmatdef2}
\end{align}
where
\begin{align}
 {\cal N}^{(\alpha)} &\equiv \text{max}_{t,t'} \int d^3r \big(\alpha^{(1)}(\rvec,t) - \alpha^{(1)}(\rvec,t')\big)^2 \\
  {\cal N}^{(\phi)} &\equiv\text{max}_{t,t'}\int d^3r \big(\boldsymbol{\phi}^{(1)}(\rvec,t) - \boldsymbol{\phi}^{(1)}(\rvec,t')\big)^2
\end{align}
are chosen to ensure normalization. This way, $R(t,t')$ can be interpreted as degree of recurrence \textit{relative} to the most extreme distinctness of occurring states. Figures~\ref{fig:recMat_alpha} and \ref{fig:recMat_phi} show graphical representations \myHighlight{called recurrence plots corresponding to} the above mentioned bubble column's recurrence matrices
\begin{equation}
 R_{m,n}\equiv R(m\Delta t_{\text{rec}},n\Delta t_{\text{rec}})\myDelete{,}\myHighlight{.}
 \label{eq:recmatdef3}
\end{equation}
\myDelete{its recurrence plots. }In this case, the choice of norm~\eqref{eq:recmatdef1} or \eqref{eq:recmatdef2} has hardly any impact. Only the anti-diagonal lines are marginally stronger with norm~\eqref{eq:recmatdef1} because it does not distinguish between motion in different directions starting from the same configuration. If necessary, this deficit could be cured by increasing the embedding dimension (\cite{Iwanski1998,March2005,Marwan2007}), i.e.\ by not only comparing single but sequences of snapshots.

Definitions~\eqref{eq:recmatdef1}, \eqref{eq:recmatdef2} and \eqref{eq:recmatdef3} generalize the original (\cite{Eckmann1987}), most widespread \textit{binary} or \textit{thresholded} recurrence matrices (i.e. $R_{m,n} \in \{0,1\}$) in two ways: Firstly, they allow for continuous extents of similarity, which is referred to as \textit{unthresholded} recurrence matrices (\cite{Marwan2007}). 
Secondly, they contain information of the whole, many-degree-of-freedom flow fields at each time step in contrast to signals from single probe points (\cite{Babaei2012,Chunhua2010,Llop2015,Sedighikamal2013,Tahmasebpour2013,Wang2012}).

\subsection{Recurrence Process}
\label{sec:recproc}
\begin{figure}[t]
\centering
  \includegraphics[height=0.3\textwidth]{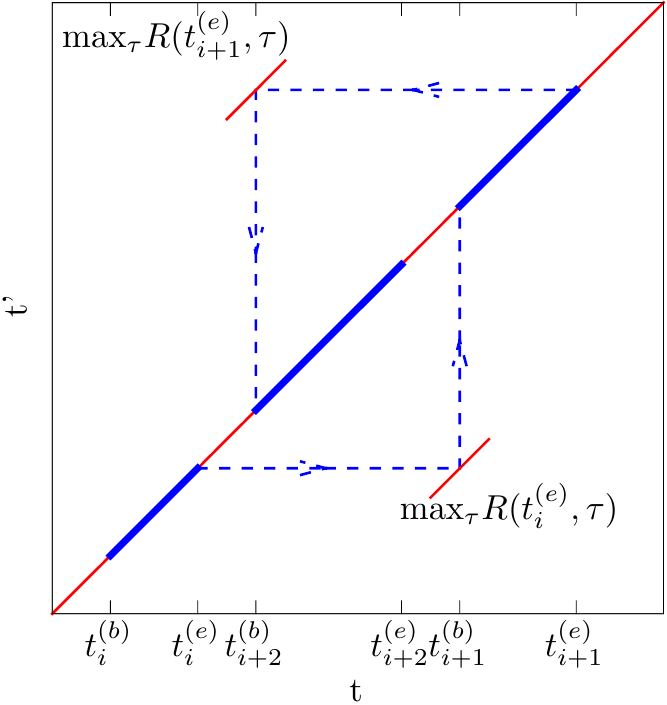}
\caption{Sketch of a reconstructed path $\{t^{\text{(b)}}_i, \dots t^{\text{(e)}}_i \sim t^{\text{(b)}}_{i+1},\dots t^{\text{(e)}}_{i+1} \sim t^{\text{(b)}}_{i+2},\dots t^{\text{(e)}}_{i+2} \dots \}$ (blue) through the recurrence plot (red). Per construction, $R(t^{\text{(e)}}_i,t^{\text{(b)}}_{i+1})$ is maximal at connections $t^{\text{(e)}}_i \sim t^{\text{(b)}}_{i+1}$. For the sake of clarity, the recurrence plot is reduced to its main diagonal and two off-diagonal maxima.}
\label{fig:recPath}
\end{figure}
Until now, recurrence matrices and their graphical representation, recurrence plots, have been used to analyze the behavior of dynamic systems. Properties like the density of recurrence points or the length distribution of diagonal lines were defined by \cite{Zbilut1992} to quantify recurrence analysis (RQA). For example, the average diagonal length, corresponds to the duration for which two parts of a system's trajectory remain close to each other given similar initial conditions at two different times and has been interpreted as mean prediction time (\cite{Marwan2002,Webber1994,Zbilut1992}).

While such measures can be used to characterize different regimes, for instance in fluidized beds (\cite{Babaei2012,Chunhua2010,Llop2015,Sedighikamal2013,Tahmasebpour2013,Wang2012}), we believe that in using recurrence statistics exclusively for data analysis, one does not exploit its full potential yet. Given a large enough recurrence matrix, it should be possible not only to reconstruct the signal from which it was created (\cite{Thiel2004}) but to extend it in a physically meaningful way far beyond recording time.

The strategy we use in this study is illustrated in Fig.~\ref{fig:recPath}. Although it does not generally produce the path that globally minimizes the induced error, it leads to very good results for sufficiently recurrent systems in a simple and fast fashion by putting \myDelete{emphasize}\myHighlight{emphasis} on connecting similar states without spurious periodicities or running into the end of the recurrence plot. Starting at some begin time $t^{\text{(b)}}_i$, we randomly pick an interval length $\Delta t_i > \Delta t_{\text{rec}} $ in a previously specified range well below the recurrence matrix size. For this interval we use the corresponding original flow data to start the reconstruction process. Then we look for a state similar to that at the interval end time $t^{\text{(e)}}_i\equiv t^{\text{(b)}}_i +\Delta t_i$ in the recurrence statistics. Depending on \myDelete{wether}\myHighlight{whether} $t_i^{\text{(e)}}$ is in the first or second half of it, we choose the next state in the other half. There, we pick $t^{\text{(b)}}_{i+1}$ such that $R(t_i^{\text{(e)}},t^{\text{(b)}}_{i+1})$ is maximized. With another randomly chosen interval length $\Delta t_{i+1}$, we can extend the constructed sequence with fields from this interval and so on. Mathematically speaking, we approximate fields with
\begin{align}
 \alpha(\rvec,t)\approx \alpha_{\text{rec}}(\rvec,t)\equiv \sum_j& \alpha\big(\rvec,t-t^{\text{(elap)}}_j+t^{\text{(b)}}_j\big)\nonumber\\
 &\Theta\big(t-t^{\text{(elap)}}_j\big)\Theta\big(t^{\text{(elap)}}_{j+1}-t\big),
 \label{eq:atilde}
\end{align}
where
\begin{equation}
 t^{\text{(elap)}}_j\equiv \sum_{i<j}t^{\text{(e)}}_i-t^{\text{(b)}}_i
\end{equation}
is the elapsed time in the recurrence process. The corresponding discrete time series is given by
\begin{align}
 \Big(\alpha_{\text{rec}}(\rvec,t^{\text{(b)}}_i)&,\alpha_{\text{rec}}(\rvec,t^{\text{(b)}}_i+\Delta t_{\text{rec}}),\dots,
 \alpha_{\text{rec}}(\rvec,t^{\text{(e)}}_i)\approx \nonumber\\
 &\alpha_{\text{rec}}(\rvec,t^{\text{(b)}}_{i+1}),\alpha_{\text{rec}}(\rvec,t^{\text{(b)}}_{i+1}+\Delta t_{\text{rec}}),\dots \Big).
  \label{eq:atildedisc}
\end{align}
Analogous expressions to Eq.~\eqref{eq:atilde} and \eqref{eq:atildedisc} are used for other fields of interest.

This procedure can be repeated to obtain an arbitrarily long sequence of generic flow patterns. Obviously, there are jumps in going from $t^{\text{(e)}}_i$ to $t^{\text{(b)}}_{i+1}$. However, based on the examples shown below, we believe that these discontinuities play a minor role for passive processes taking place on the flow for the case of a suitably chosen recurrence matrix.

\subsection{Recurrence CFD}
In the following, we present two strategies to simulate passive processes on recurrence fields, either in a Eulerian (Model A) or Lagrangian (Model B) fashion. For the sake of simplicity, we restrict ourselves to incompressible fluids, but we stress that the generalization to the compressible case is straight-forward.
\subsubsection*{Model A: Passive Transport Equation}
We solve a transport equation
\begin{align}
 \frac{\partial }{\partial t}\alpha_{\text{rec}}&(\rvec,t)c(\rvec,t)+\nabla\cdot \boldsymbol{\phi}_{\text{rec}}(\rvec,t)c(\rvec,t) \nonumber\\
 &- \nabla \cdot  \alpha_{\text{rec}}(\rvec,t)D(\rvec)\nabla c(\rvec,t)= S(\rvec,t)
 \label{eq:transeq1}
\end{align}
for the convection and diffusion with diffusion coefficient $D(\rvec)$ of a passive species concentration $c(\rvec,t)$ on the recurrence fields of one of the phases. For the sake of brevity, the phase index is omitted. Employing definition~\eqref{eq:atilde} for the approximation of the volume fraction and the flux, it is straight-forward to show that
% \begin{alignat}{2}
%  & \frac{\partial }{\partial t}&&\alpha_{\text{rec}}(\rvec,t)c(\rvec,t)+\nabla\cdot \boldsymbol{\phi}_{\text{rec}}(\rvec,t)c(\rvec,t)\nonumber\\
%  & &&- \nabla \cdot \alpha_{\text{rec}}(\rvec,t)D(\rvec)\nabla c(\rvec,t)= \nonumber\\
%  & \frac{\partial }{\partial t}&&\alpha(\rvec,t)c(\rvec,t)+\nabla\cdot \boldsymbol{\phi}(\rvec,t)c(\rvec,t) \nonumber\\
%  & &&- \nabla \cdot  \alpha(\rvec,t)D(\rvec)\nabla c(\rvec,t) 
%  -  S^{\text{(rec)}}(\rvec,t),
% \end{alignat}
\begin{align}
  &\frac{\partial }{\partial t}\alpha_{\text{rec}}(\rvec,t)c(\rvec,t)+\nabla\cdot \boldsymbol{\phi}_{\text{rec}}(\rvec,t)c(\rvec,t)\nonumber\\
 &\qquad\qquad\qquad - \nabla \cdot \alpha_{\text{rec}}(\rvec,t)D(\rvec)\nabla c(\rvec,t)= \nonumber\\
 & \frac{\partial }{\partial t}\alpha(\rvec,t)c(\rvec,t)+\nabla\cdot \boldsymbol{\phi}(\rvec,t)c(\rvec,t) \nonumber\\
 &\qquad\qquad\qquad - \nabla \cdot  \alpha(\rvec,t)D(\rvec)\nabla c(\rvec,t) 
 -  S^{\text{(rec)}}(\rvec,t),
\end{align}
where the recurrence-error source
\begin{align}
S^{\text{(rec)}}&(\rvec,t) \equiv \nonumber\\
&S^{\text{(rec)}}_{\text{t-jmp}}(\rvec,t) +S^{\text{(rec)}}_{\text{t}}(\rvec,t)+S^{\text{(rec)}}_{\text{conv}}(\rvec,t)+S^{\text{(rec)}}_{\text{diff}}(\rvec,t) 
\end{align}
contains contributions from the time-derivative, the convective and the diffusive terms,
\begin{align}
 S&^{\text{(rec)}}_{\text{t-jmp}}\big(\rvec,t\big) \equiv\nonumber \\
 &\qquad\sum_j \delta\big(t-t^{\text{(elap)}}_j\big)\big(\alpha(\rvec,t^{\text{(e)}}_{j-1})-\alpha(\rvec,t^{\text{(b)}}_{j})\big)c(\rvec,t)  \label{eq:serr1}\\
 S&^{\text{(rec)}}_{t}\big(\rvec,t^{\text{(elap)}}_j\leq t<t^{\text{(elap)}}_{j+1}\big) \equiv \nonumber \\
 &\qquad
 \frac{\partial }{\partial t}\big(\alpha(\rvec,t)-\alpha(\rvec,t-t^{\text{(elap)}}_j+t^{\text{(b)}}_j)\big)c(\rvec,t) \label{eq:serr2}\\
 S&^{\text{(rec)}}_{\text{conv}}\big(\rvec,t^{\text{(elap)}}_j\leq t<t^{\text{(elap)}}_{j+1}\big) \equiv \nonumber\\
 &\qquad \nabla\cdot \big(\boldsymbol{\phi}(\rvec,t)-\boldsymbol{\phi}(\rvec,t-t^{\text{(elap)}}_j+t^{\text{(b)}}_j)\big)c(\rvec,t) \label{eq:serr3}\\
 S&^{\text{(rec)}}_{\text{diff}}\big(\rvec,t^{\text{(elap)}}_j\leq t<t^{\text{(elap)}}_{j+1}\big) \equiv\nonumber\\
 &\qquad-\nabla\cdot  \big(\alpha(\rvec,t)-\alpha(\rvec,t-t^{\text{(elap)}}_j+t^{\text{(b)}}_j)\big)D(\rvec)\nabla c(\rvec,t).
 \label{eq:serr4}
\end{align}
Solving a transport equation on recurrence fields therefore imposes errors in form of additional, artificial source terms coupling to the concentration field and its temporal and spatial derivatives. This means that in comparison to the solution on the genuine fields, Eq.~\eqref{eq:transeq1} corresponds to the temporally highly non-local
\begin{align}
 \frac{\partial }{\partial t}\alpha(\rvec,t)c(\rvec,t)+&\nabla\cdot \boldsymbol{\phi}(\rvec,t)c(\rvec,t) - \nabla \cdot  \alpha(t)D\nabla c(t) = \nonumber\\
 &S(\rvec,t)+ S^{\text{(rec)}}(\myHighlight{\rvec, }t).
  \label{eq:transeq2}
\end{align}
Among the various contributions Eqs.~\eqref{eq:serr1} -- \eqref{eq:serr4} to $S^{\text{(rec)}}(\myHighlight{\rvec, }t)$, $ S^{\text{(rec)}}_{\text{t-jmp}}(\rvec,t)$ differs from the others by acting in a discrete way only at jumps in the recurrence statistics while the others describe a continuous, possible long-term divergence of trajectories. In practice, these are not directly accessible without knowledge of the fields at all times whereas $S^{\text{(rec)}}_{\text{t-jmp}}(\rvec,t)$ requires only data from the recording period. Its effect can be mitigated both by increasing the size of the recurrence matrix and correspondingly that of the time intervals $\Delta t_i$ and by choosing appropriate $\big(t^{\text{(e)}}_{j-1},t^{\text{(b)}}_{j}\big)$ pairs. Assuming a truly recurrent system whose degree of self-similarity does not significantly fade after several pseudo-periods, one can argue by induction that $S^{\text{(rec)}}_{\text{t}}(\rvec,t)$ and $S^{\text{(rec)}}_{\text{diff}}(\rvec,t)$ are negligible if $\alpha(\myHighlight{\rvec, }t^{\text{(e)}}_{j-1})\sim\alpha(\myHighlight{\rvec, }t^{\text{(b)}}_{j})$ for all $j$. \myDelete{Somewhat handwavingly, we assume that}\myHighlight{Since $\alpha(\rvec,t)$ and $\boldsymbol{\phi}(\rvec,t)$ are connected via the continuity equation,} a similar evolution of volume fraction fields indicates the same for \myHighlight{the irrotational parts of} the corresponding fluxes so that $S^{\text{(rec)}}_{\text{conv}}(\rvec,t)$ \myDelete{is}\myHighlight{might be} small, too. This reasoning supports $R^{(\alpha)}(t,t')$ as a meaningful starting base to approximate the genuine transport equation with Eq.~\eqref{eq:transeq1}, but 
of course one could also make the reverse argument for $R^{(\phi)}(t,t')$\myDelete{.}\myHighlight{: similar fluxes imply similar volume fractions up to a constant.}

\myHighlight{We stress, however, that these considerations need to be taken with caution because they are not unconditionally valid, e.g.\ for the former case in the presence of dominating vortices. This underlines our opinion that an appropriate recurrence norm has to be chosen carefully for each system as part of the modelling.}

\myHighlight{If $R^{(\alpha)}(t,t')$ turns out to be a meaningful option,} additional support \myDelete{for the former choice} within Model A comes from an analysis of the errors imposed by approximations of the form $\alpha(\rvec,t)\approx \alpha_{\text{rec}}(\rvec,t)$ within Eq.~\eqref{eq:transeq1}. While \textit{global} conservation of species without transport across domain boundaries,
\begin{equation}
 \frac{\partial C(t)}{\partial t}\equiv \frac{\partial}{\partial t} \int d^3r\alpha_{\text{rec}}(\rvec,t)c(\rvec,t) = \int d^3r S(\rvec,t),
 \end{equation}
immediately follows from volume integration of Eq.~\eqref{eq:transeq1} and application of Gauss' law, \textit{local} discontinuities
\begin{equation}
 c(\rvec,t\searrow t^{\text{(elap)}}_j) = \frac{\alpha(\rvec,t^{\text{(e)}}_{j-1})}{\alpha(\rvec,t^{\text{(b)}}_j)}c(\rvec,t\nearrow t^{\text{(elap)}}_j),
 \label{eq:locdisc}
\end{equation}
where $\searrow$ and $\nearrow$ mean limits from above and blow, occur at jumps of $\alpha_{\text{rec}}(\rvec,t)$ in the recurrence statistics, but can be minimized along a path constructed from $R^{(\alpha)}(t,t')$. Equally present jumps of $\boldsymbol{\phi}_{\text{rec}}(\rvec,t)$ can impair the smoothness of the evolution of $c(\rvec,t)$, but do not affect continuity and local conservation of species, respectively. Physically speaking, fluxes do not create or remove mass but transport it from one region to another.

In the worst case, Eq.~\eqref{eq:locdisc} \myDelete{leads to divergencies}\myHighlight{diverges} when $\alpha(\rvec,t^{\text{(b)}}_j)=0$ but $c(\rvec,t\nearrow t^{\text{(elap)}}_j) \neq 0$, i.e.\ if at some point no material is present to carry a finite amount of the passive species. Such cases need to be taken care of, for example by redistributing the excess of species.

Having decided on a specific recurrence norm, $\alpha_{\text{rec}}(\rvec,t)$ and $\boldsymbol{\phi}_{\text{rec}}(\rvec,t)$ can be obtained and Eq.~\eqref{eq:transeq2} solved, for example with a finite-volume approach for spatial and any marching method for temporal discretization. In dependence on a typical discretization length $\Delta l$ and flow velocity $u$, we choose the computational time step $\Delta t$ such that the Courant number satisfies
\begin{equation}
 \textsf{Co} = \frac{u\Delta t}{\Delta l} \lesssim 1.
 \label{eq:tco}
\end{equation}
 The resulting computational time step might be much smaller than the monitoring time step $\Delta t_{\text{rec}}$. However, no additional information about $\alpha_{\text{rec}}(\rvec,t)$ and $\boldsymbol{\phi}_{\text{rec}}(\rvec,t)$ between monitoring times is needed because the choice Eq.~\eqref{eq:tref} for $\Delta t_{\text{rec}}$ ensures that the fields at monitoring times are good approximations to those in-between.

\subsubsection*{Model B: Lagrangian (Fluid) Particles}
Regardless if we consider a fluid from a Lagrangian perspective, where we first have to decompose it into parcels of equal mass, i.e.\ portions of coherent fluid volumes $V_{\text{p}}$, or actual physical particles (two terms we will use synonymously in the following), they are connected to field quantities via some averaging function $g(r)$ with $\int d^3r g(r) =1$ (\cite{AndersonJackson1967}). For the sake of simplicity, we assume incompressible conditions and define
\begin{align}
 \alpha(\rvec,t)&=V_{\text{p}}\sum_i g(|\rvec-\rvec_i|) \label{eq:alphapart}\\
 \uvec(\rvec,t)&=\frac{\sum_i g(|\rvec-\rvec_i|)\uvec_i}{\sum_i g(|\rvec-\rvec_i|)}.
\end{align}
The particles' equation of motion are modelled with
\begin{equation}
d\rvec_i=\uvec_{\text{rec}}(\rvec_i,t)dt + \sqrt{2D_{\text{rec}}(\rvec_i,t)}d\textbf{w}_i.\label{eq:newton2}
\end{equation}
The first term on the right-hand side corresponds to convection with the mean velocity obtained from the recurrence statistics at the current particle position. The second one allows for independent, normal-distributed displacements $d\textbf{w}_{i;x,y,z}\sim {\cal N}(0,dt)$, i.e.\ fluctuations leading to Brownian motion with a yet undefined diffusion coefficient $D_{\text{rec}}(\rvec_i,t)$.
Since the mean-square displacement due to fluctuations,
\begin{equation}
 dr_i^2=2D_{\text{rec}} d\textbf{w}_i^2=6D_{\text{rec}} dt
\end{equation}
grows linearly with time, the expansion of a sufficiently differentiable function of $\rvec_i$ to first order in $dt$ needs to also include terms of second order in $d\textbf{w}_i$ as formalized by It\^{o}'s lemma (\cite{Oksendal2003}). Specifically for $g(|\rvec-\rvec_i|)$, one finds
\begin{align}
 dg(|\rvec&-\rvec_i|) \approx d\rvec_i\cdot\nabla_i g(|\rvec-\rvec_i|) + \frac{1}{2}\big(d\rvec_i\cdot\nabla_i\big)^2 g(|\rvec-\rvec_i|)\nonumber\\
 &\approx\big(\uvec_{\text{rec}}(\rvec_i,t)dt + \sqrt{2D_{\text{rec}}(\rvec_i,t)}d\textbf{w}_i\big)\cdot\nabla_i g(|\rvec-\rvec_i|)\nonumber\\
 & + D_{\text{rec}}(\rvec_i,t)\big(d\textbf{w}_i\cdot\nabla_i\big)^2 g(|\rvec-\rvec_i|),
\end{align}
where $\nabla_i\equiv \partial / \partial \rvec_i$.
For isotropic fluctuations, the linear term in $d\textbf{w}_i$ vanishes and one recovers the Feynman-Kac formula (\cite{Oksendal2003})
\begin{align}
 \frac{\partial g(|\rvec-\rvec_i|)}{\partial t} = \uvec_{\text{rec}}(\rvec_i,t)\cdot&\nabla_i g(|\rvec-\rvec_i|) \nonumber\\
 &+ D_{\text{rec}}(\rvec_i,t)\nabla_i^2 g(|\rvec-\rvec_i|).\label{eq:g1}
\end{align}
By connecting the diffusion coefficient to the excess in volume fraction, ${\delta\alpha(\rvec,t)\equiv \text{max}\Big[\alpha(\rvec,t) - \alpha^{\text{(rec)}}(\rvec,t),0\Big]}$, e.g.\ via
\begin{equation}
 D_{\text{rec}}(\rvec,t)=D_{\text{rec}}^{(0)}\frac{\delta\alpha(\rvec,t)}{\alpha(\rvec,t)},
\end{equation}
it is possible to mitigate deviations of the particle position field Eq.~\eqref{eq:alphapart} from $\alpha_{\text{rec}}(\rvec,t)$ due to numerical errors or jumps in the recurrence statistics.
Summing over all particles and using $\nabla_i g(|\rvec-\rvec_i|)=-\nabla g(|\rvec-\rvec_i|)$, this gives
\begin{align}
   \frac{\partial}{\partial t}\alpha(\rvec,t)+\nabla\cdot \alpha(\rvec,t)&\uvec_{\text{rec}}(\rvec,t) = \nonumber\\
   &\nabla\cdot D^{(0)}_{\text{rec}} \nabla\delta\alpha(\rvec,t)+{\cal O}(\nabla^3),\label{eq:contarelax}
\end{align}
which is nothing else than the well-known continuity equation plus a relaxation term in form of a diffusive flux $\textbf{j}_{\text{diff}}(\rvec,t)\equiv D^{(0)}_{\text{rec}} \nabla\delta\alpha(\rvec,t)$ from regions containing too many particles in comparison to the recurrence field. 
This is of course a less brutal and by definition mass-conserving way to force the current volume fraction field to the value obtained from the recurrence statistics than by simply resetting it as done in Model A. In particular, the extent of forcing can be controlled via $D^{(0)}_{\text{rec}}$\myDelete{, $l_{\text{f}}$ and $\tau_{\text{f}}$, respectively}.

At this point, we want to stress that we do \textit{not} solve Eq.~\eqref{eq:contarelax}, but use it to interpret the relationship of Eq.~\eqref{eq:newton2} from which it is derived, with Eq.~\eqref{eq:transeq1} of Model A. Instead, we integrate Eq.~\eqref{eq:newton2}, where we employ Donsker's extension of the central limit theorem (\cite{Donsker1951,Fischer2010}), which allows us to approximate Brownian motion with a random walk. Since such a walk's mean-square displacement after $n$ steps is given by $\langle r^2(n)\rangle=nl_{\text{f}}^2$, we relate its step size $l_{\text{f}}(\rvec_i,t)$ and time $\tau_{\text{f}}$ to $D_{\text{rec}}$ by
\begin{equation}
 6D_{\text{rec}}(\rvec_i,t)=\frac{l_{\text{f}}^2(\rvec_i,t)}{\tau_{\text{f}}}.
\end{equation}
In contrast to Model A, Model B does not exhibit artificial local creation or destruction of mass because the particles conserve it by definition. To describe the same transport processes as in Model A, at least two scenarios are possible: Either let each particle $i$ explicitly carry some concentration $c_i(t)$ which determines the field value via
\begin{equation}
c(\rvec,t)=\frac{\sum_i g(|\rvec-\rvec_i|)c_i(t)}{\sum_i g(|\rvec-\rvec_i|)},
\end{equation}
or solve a transport equation \eqref{eq:transeq1} using $\alpha(\rvec,t)$ obtained from the current particle positions according to Eq.~\eqref{eq:newton2}.
Although both come at the cost of additionally propagating particle positions, it is worth noting that this can be done in a local and explicit manner, which paves the ground for highly efficient parallelisation on distributed compute cores.

Since the second of the above approaches is more or less a combination of the first one and Model A, we focus here on the former.
It automatically describes convective species transport, but needs additional care to handle diffusion as becomes apparent by multiplying Eq.~\eqref{eq:g1} with $c_i(t)$ before summing over all particles.
Arguing along the same lines as above, one gets
\begin{align}
   \frac{\partial}{\partial t}\alpha(\rvec,t)&c(\rvec,t)+\nabla\cdot \alpha(\rvec,t)\uvec_{\text{rec}}(\rvec,t)c(\rvec,t) = \nonumber\\
   &\nabla\cdot D^{(0)}_{\text{rec}} \nabla\delta\alpha(\rvec,t)c(\rvec,t)+\sum_ig(|\rvec-\rvec_i|) \dot{c}_i,\label{eq:transeqB}
\end{align}
where the explicit time dependence of each particle's species concentration
\begin{equation}
 \dot{c}_i(t)\equiv\dot{c}_i^{\text{(source/sink)}}(t)+\dot{c}_i^{\text{(diff)}}(t)
\end{equation}
is due to sources or sinks and diffusion processes between particles. Besides physical diffusion which could be modelled for example by allowing for species transfer via contact detection or the introduction of a relaxation mechanism towards the local average, Eq.~\eqref{eq:transeqB} contains an artificial diffusion-like term $\nabla\cdot D^{(0)}_{\text{rec}} \nabla\delta\alpha(\rvec,t)c(\rvec,t)$. It describes transport with fluctuations and depending on the specific system and effects of interest, it might exceed physical diffusion and necessitate further modelling.

The numerical treatment of Model B comprises choices of an integrator for the particle trajectories, time steps for recurrence fields and particle position updates and a realisation of the fluctuations. As mentioned above, the latter can be approximated with a random walk, e.g.\ with isotropically distributed displacements with step size $l_{\text{f}}$ and step time $\tau_{\text{f}}$ which may be reasonably identified with the integration time step $\Delta t$.
One can then propagate particle positions according to a discretized form of Eq.~\eqref{eq:newton2}, most simply the explicit first-order Euler-Maruyama scheme (\cite{KloedenPlaten}), the adaption of the well-known Euler method to stochastic differential equations.

To deduce criteria for the time steps involved, we estimate the displacement of a particle due to convection between two recurrence frames,
\begin{align}
 &\rvec(t+\Delta t_{\text{rec}})-\rvec(t)=\nonumber\\
 &\qquad\int_t^{t+\Delta t_{\text{rec}}}dt' \uvec_{\text{rec}}\big(\rvec(t'),t'\big)= \nonumber \\
 &\qquad\int_t^{t+\Delta t_{\text{rec}}}dt'\Big[\uvec_{\text{rec}}\big(\rvec(t'),t\big)\nonumber\\
 &\qquad\qquad\qquad\qquad+(t'-t)\frac{\partial}{\partial t}\uvec_{\text{rec}}\big(\rvec(t'),t\big)+\dots\Big]\approx\nonumber\\
 &\qquad \int_t^{t+\Delta t_{\text{rec}}}dt'\uvec_{\text{rec}}\big(\rvec(t'),t\big).\label{eq:disp1}
\end{align}
The last approximation holds if
\begin{equation}
 |\uvec_{\text{rec}}(\rvec,t)| \gg \Delta t_{\text{rec}}\Big|\frac{\partial}{\partial t}\uvec_{\text{rec}}(\rvec,t)\Big|,\label{eq:trecB}
\end{equation}
which is just the criterion stated in Eq.~\eqref{eq:tref0}. Then it is legitimate to update recurrence fields not before $\Delta t_{\text{rec}}$. To further evaluate Eq.~\eqref{eq:disp1}, $(t,t+\Delta t_{\text{rec}})$ is partitioned into intervals of the computation time step size,
\begin{align}
 &\int_t^{t+\Delta t_{\text{rec}}}dt'\uvec_{\text{rec}}\big(\rvec(t'),t\big) = \nonumber\\
 &\qquad\sum_{i} \int_{t+i\Delta t}^{t+(i+1)\Delta t}dt'\uvec_{\text{rec}}\big(\rvec(t'),t\big)\approx \nonumber \\
 &\qquad\sum_{i} \int_{t+i\Delta t}^{t+(i+1)\Delta t}dt'\uvec_{\text{rec}}\big(\rvec(t+i\Delta t),t\big)=\nonumber\\
 &\qquad\sum_{i} \Delta t\uvec_{\text{rec}}\big(\rvec(t+i\Delta t),t\big),\label{eq:eulerapprox2}
\end{align}
which resembles the desired first-order Euler method. Arguing along the same lines as above, the approximation in Eq.~\eqref{eq:eulerapprox2} is valid if
\begin{equation}
 |\uvec_{\text{rec}}(\rvec,t)|\gg \Delta t | \uvec_{\text{rec}}(\rvec,t)\cdot\nabla\uvec_{\text{rec}}(\rvec,t)|.\label{eq:dtB}
\end{equation}
We interpret both conditions Eqs.~\eqref{eq:trecB} and \eqref{eq:dtB} in the average sense of Eq.~\eqref{eq:tref} to obtain meaningful values for $\Delta t_{\text{rec}}$ and $\Delta t$.

% ----------------------------------------------------------------------
\section{Results}
We implemented our recurrence CFD models into the prominent open-source CFD-DEM code CFDEMcoupling (\cite{Goniva2012})
and into the commercial code 
ANSYS/Fluent. In the former case, specialized classes and solvers were written \myHighlight{and applied to the example of a bubble column (Sec.~\ref{sec:obc})}, in the latter one, we 
realized the complete recurrence CFD procedure with help of user defined functions (UDFs) \myHighlight{and tested them on simulations of steel decarbonisation (Sec.~\ref{sec:bof})}.

\subsection{Oscillating Bubble Column}
\label{sec:obc}

% motivate the osc.bub.col. simulations
As a first test case we considered an oscillating air-water bubble column. Despite their seemingly conceptual simplicity, such systems have inspired a huge amount of both experimental and computational research, see for example the references listed by \cite{Tabib2008}. These efforts have partly been driven by the industrial need to understand bubble-column reactors exhibiting a large variety of both length and time scales. As stated by \cite{Becker1999}, turbulent fluctuations are in the range of milliseconds and below while mixing due to large-scale circulation takes places within seconds or minutes and material changes due to chemical reactions happen over hours. In this sense, bubble columns may be seen as prototype of system suitable for the description with recurrence-based models. Indeed, other statistical methods have already been used to characterize flow transitions, cf.\ \cite{Drahovs1991,Letzel1997,Lin2001}.

Although CFD simulations can reproduce the experimentally observed oscillation frequency very well (\cite{Becker1994,Sokolichin1994}),
we do not focus on their validity here. Rather, we compare the 
prediction of our recurrence formalism with the numerical experiment of a full CFD simulation which 
we consider as a valid base.
\subsubsection{Simulation set-up}

% geometry
\begin{figure}[htbp]
	\centering
	\small
	\includegraphics[width=0.4\textwidth]{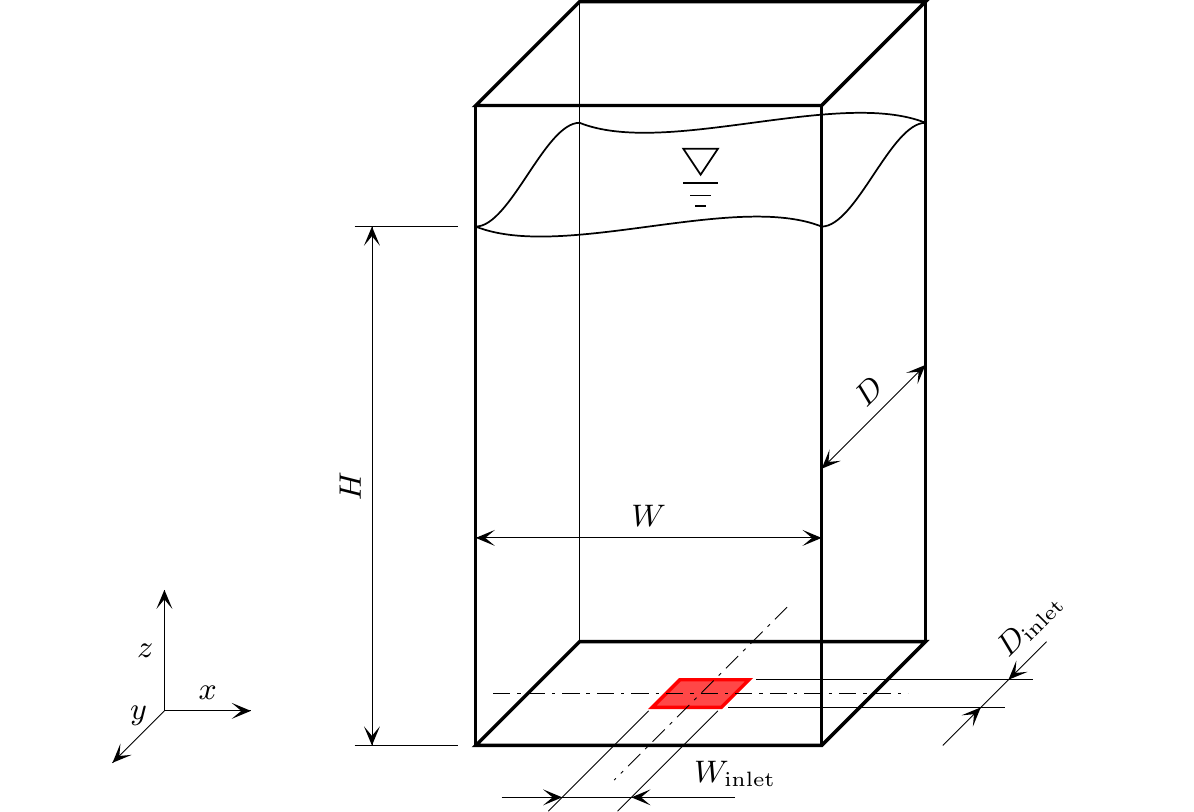}
	\caption{Geometry of the centrally aerated bubble column. The colored patch 
		marks the gas inlet, the water surface is modelled as outlet (see Tab.~\ref{tab:beckerBubbleColumnBC}). The dimensions can be found in Tab.~\ref{tab:beckerBubbleColumnDimensions}.
	}
	\label{fig:beckerBubbleColumnGeometry}
\end{figure}

% dimensions
\begin{table}[htb]
	\centering
%	\begin{tabular}{lccccc}
\begin{tabular}{llllll}
	\toprule
	& $W$ & $D$ & $H$ & $W_{\mathrm{inlet}}$ & $D_{\mathrm{inlet}}$ \\
	\midrule
	length [m] & $0.2 $ & $0.05$ & $0.5$ & $0.03$ & $0.015$ \\
	number of cells & 40 & 11 & 100 & 6 & 3 \\
	\bottomrule
	\end{tabular}
	\caption{The main dimensions and number of cells in each direction (total 44000) of the oscillating bubble column geometry.}
	\label{tab:beckerBubbleColumnDimensions}
\end{table}

% BC
\begin{table*}[htb]
	\centering
	\begin{tabular}{llll}
	\toprule
	& walls & inlet & outlet \\
	\midrule
	$U_{\text{air}}$ [$\meter/\second$]& \scriptsize{\texttt{fixedValue}}, $\textbf{0}$ & \scriptsize{\texttt{fixedValue}}, $0.53\textbf{e}_{\text{z}}$ & \scriptsize{\texttt{pressureInletOutletVelocity}} \\
	$U_{\text{water}}$ [$\meter/\second$] & \scriptsize{\texttt{fixedValue}}, $\textbf{0}$ & \scriptsize{\texttt{fixedValue}}, $\textbf{0}$ &  \scriptsize{\texttt{slip}}\\
	$p$ [$\pascal$]& \scriptsize{\texttt{fixedFluxPressure}}  & \scriptsize{\texttt{fixedFluxPressure}} & \scriptsize{\texttt{fixedValue}}, $10^5$ \\
	\bottomrule
	\end{tabular}
	\caption{Types and values of boundary conditions of the oscillating bubble column in the language of OpenFOAM. \texttt{fixedValue} sets a field to a predefined value, \texttt{slip} its normal component and the gradient of its tangential components to 0. \texttt{pressureInletOutletVelocity} and \texttt{fixedFluxPressure} are both closely related to a zero-gradient condition. More specifically, \texttt{pressureInletOutletVelocity} distinguishes between in- and outflow through a surface depending on the sign of a field's flux: for inflow, the field is set to a fixed value and for outflow, its gradient to 0. \texttt{fixedFluxPressure} chooses the pressure gradient such that it agrees with the velocity field's flux on the boundary.	
	}
	\label{tab:beckerBubbleColumnBC}
\end{table*}

\begin{figure}[htbp]
	\centering
	\includegraphics[width=0.2\textwidth]{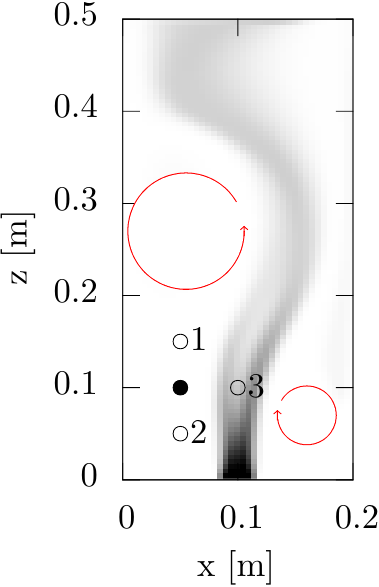}
	\caption{The locations of the injection point (full circle) and the three 
		sampling points (open circles) in the midplane. As a guide to the eye, a snapshot of the gas phase's volume fraction field 
		is shown together with indications of vortices (red) moving down on either side of the column.}
	\label{fig:pointLocations}
\end{figure}

\begin{figure*}[htbp]
\centering
\subfloat[\label{fig:sigAlpha}]{%
  \includegraphics[width=0.4\textwidth]{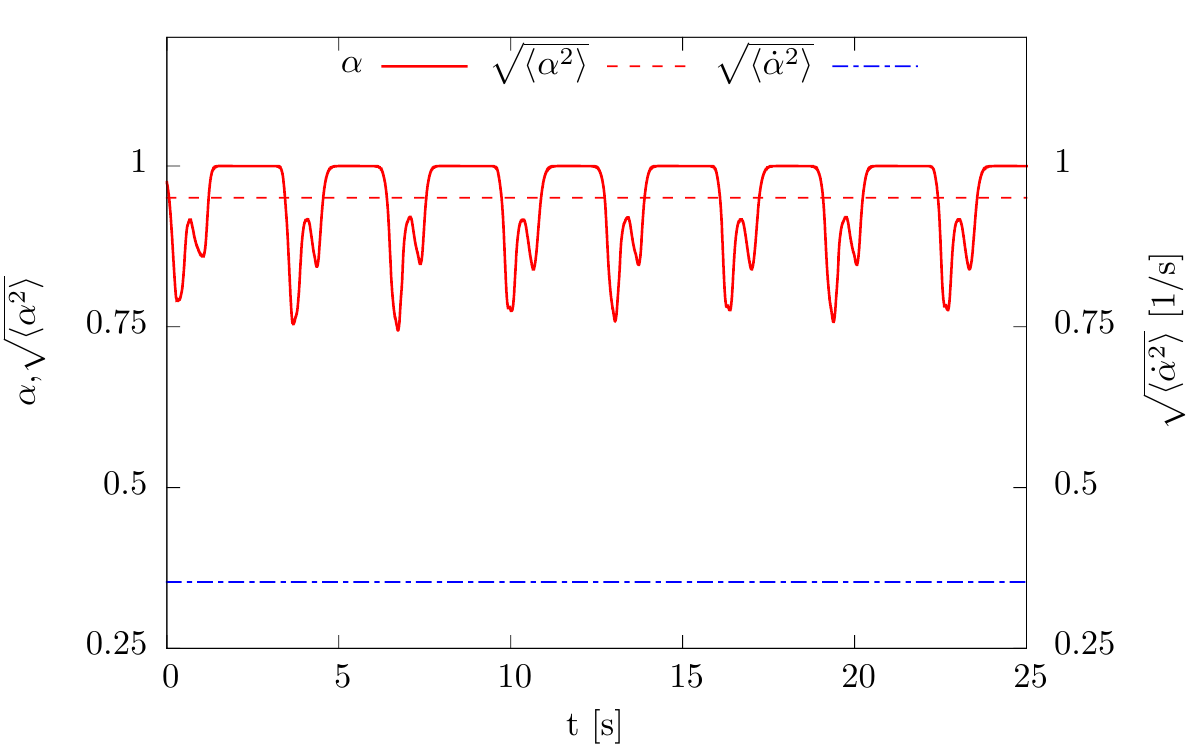}
}\hfill
\subfloat[\label{fig:sigVx}]{%
\includegraphics[width=0.4\textwidth]{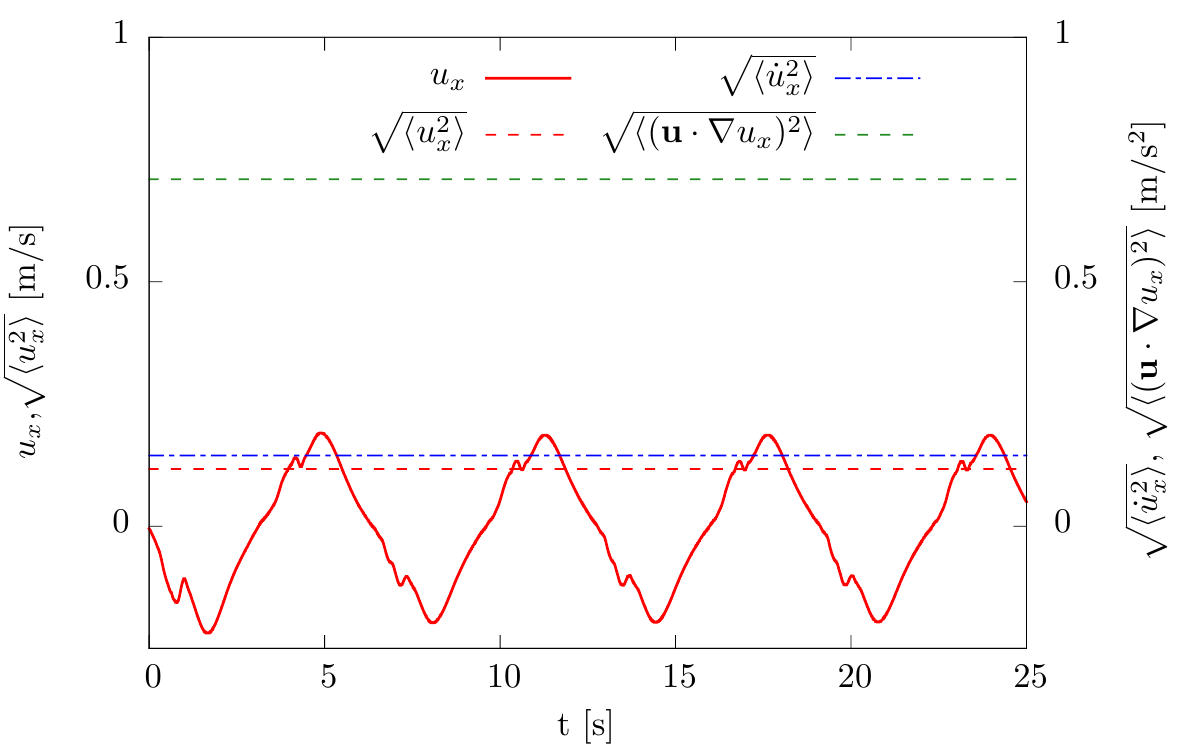}
}\\
\subfloat[\label{fig:sigVz}]{%
  \includegraphics[width=0.4\textwidth]{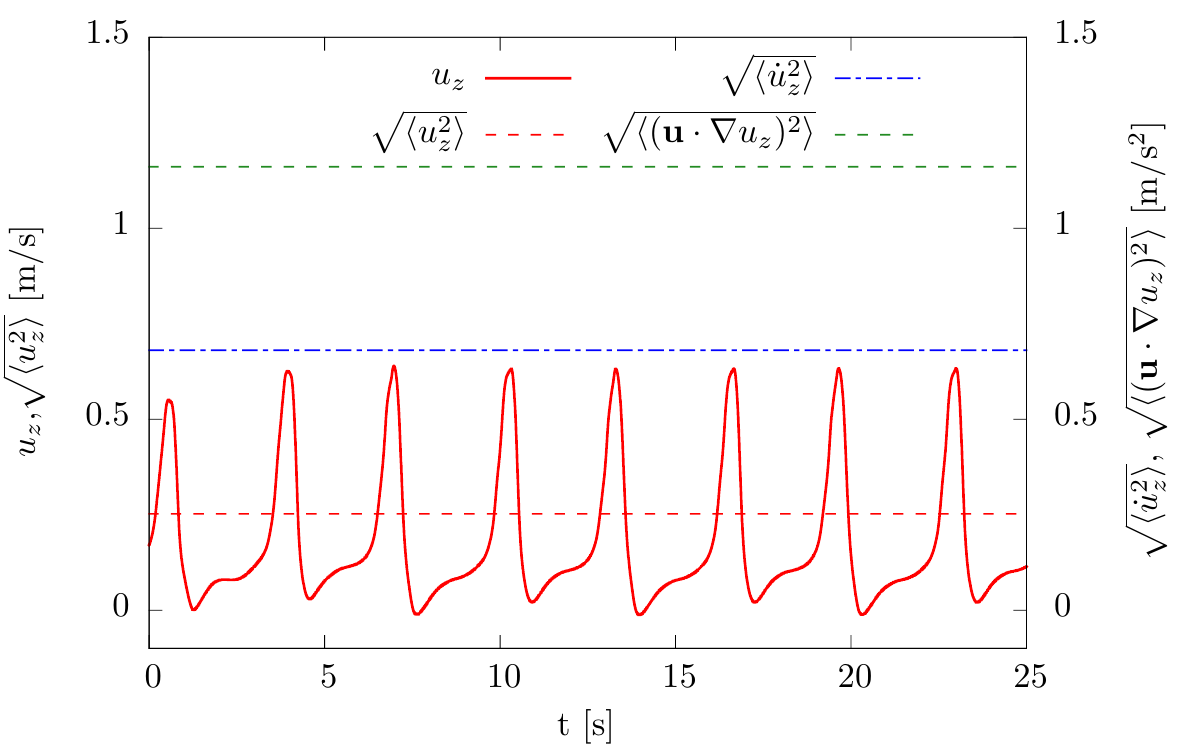}
}\hfill
\subfloat[\label{fig:sigFourier}]{%
\includegraphics[width=0.4\textwidth]{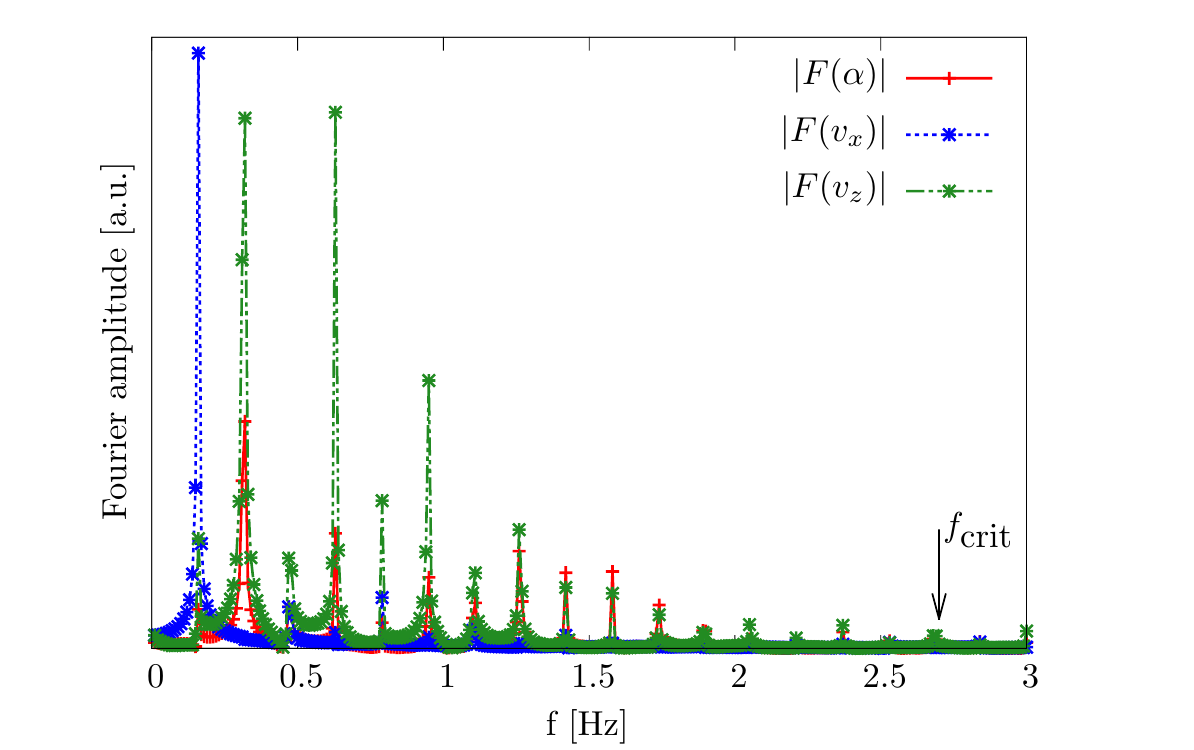}
}
	\caption{Probe values of the (a) volume fraction and (b) x- and (c) z-components of fluid velocity at sampling point 3 of Fig.~\ref{fig:pointLocations}, together with average values of the quantities, their time- and in the case of velocity also convective derivatives. Note that the actual period is best seen from the x-component (b), where in contrast to (a) and (c), the difference between back and forth movement is visible. (d) shows the absolute value of the Fourier components including the critical sampling frequency $f_{\text{crit}}$ that has to be exceeded to enable accurate interpolation of the recorded fields. The lowest-lying peak on the other hand determines the period the recurrence statistics has at least to contain.}
	\label{fig:probevaluesSignals}
\end{figure*}

% solution procedure
We chose a simple, flat geometry specified in Fig.~\ref{fig:beckerBubbleColumnGeometry} and Tabs.~\ref{tab:beckerBubbleColumnDimensions} and \ref{tab:beckerBubbleColumnBC} to simulate an oscillating bubble column with OpenFOAM's Eulerian-Eulerian two-phase solver \texttt{twoPhaseEulerFoam} by \cite{Rusche2003} which we extended with a scalar transport equation for a passive species in the liquid phase. 
After an equilibration phase during which the oscillatory state developed, the actual simulation of interest was carried out for $100 \,\second$ real time. Passive tracers in form of an additional species concentration field were injected into the liquid phase with a constant rate in a small region depicted in Fig.~\ref{fig:pointLocations} over the first $75 \,\second$, then the source was turned off. To investigate the dynamics and transport properties of the system, velocities and volume fractions of the phases as well as species concentrations were monitored at three sampling points which are also shown in Fig.~\ref{fig:pointLocations}. From the former, displayed in Fig.~\ref{fig:probevaluesSignals}, reasonable values for the monitoring time step $\Delta t_{\text{rec}}$ and for field and particle propagation time steps within Models A and B, $\Delta t_{\text{A}}$ and $\Delta t_{\text{B}}$, could be estimated according to Eqs.~\eqref{eq:tref}, \eqref{eq:tco}, \eqref{eq:trecB} and \eqref{eq:dtB}. Together with the time step $\Delta t$ of the full CFD simulation, they are shown in Tab.~\ref{tab:timeSteps}. Furthermore, a pseudo-period of $\tau_{\text{p-p}}\approx 8 \,\second$ is visible in these signals so that we constructed the recurrence statistics from the fields of the first $\tau_{\text{rec}}=25 \,\second$ after equilibration. Paths on this matrix were obtained as described in section~\ref{sec:recproc} with interval lengths in $[\tau_{\text{rec}}/20;\tau_{\text{rec}}/5]$ between jumps. With hardly any difference in the recurrence matrices shown in Figs.~\ref{fig:recMat_alpha} and \ref{fig:recMat_phi}, we used $R^{(\phi)}(t,t')$ for both models in the present study.

The setup of Model A was straight-forward in this case. The same geometry and boundary conditions as for the full simulation could be used. 
Some additional care had to be taken for Model B. In order to confine the ``water-parcels'' to the simulation box without costly particle-wall interactions, we imposed reflective boundary conditions. Although the number of particles and the magnitude of their fluctuations necessary to obtain good results are generally not known a priori and need to be determined empirically, we assumed reasonable starting values to be a few parcels per computational cell and random walk steps of about the average cell size.

\begin{figure*}[htbp]
\centering
\subfloat[\label{fig:vel_rec}]{%
   \includegraphics[width=0.25\textwidth]{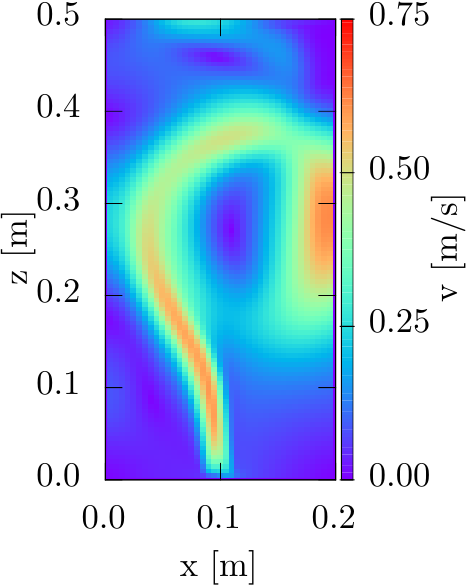}
}\hfill
\subfloat[\label{fig:vel_noFluc}]{%
 \includegraphics[width=0.25\textwidth]{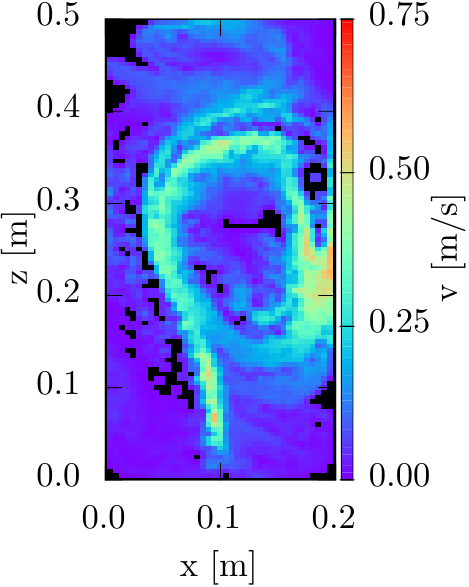}
}\hfill
\subfloat[\label{fig:vel_fluc}]{%
 \includegraphics[width=0.25\textwidth]{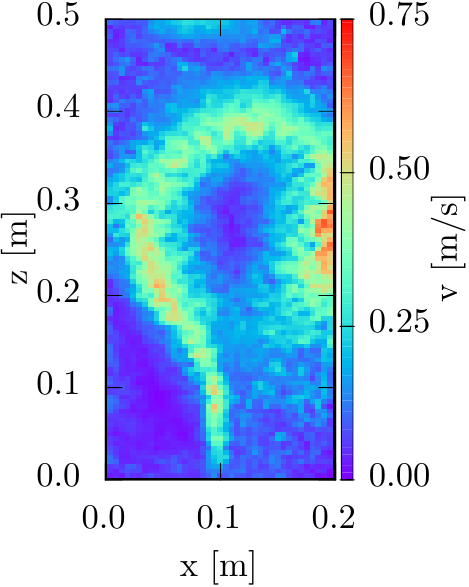}
}\\
\subfloat[\label{fig:volFrac}]{%
   \includegraphics[width=0.25\textwidth]{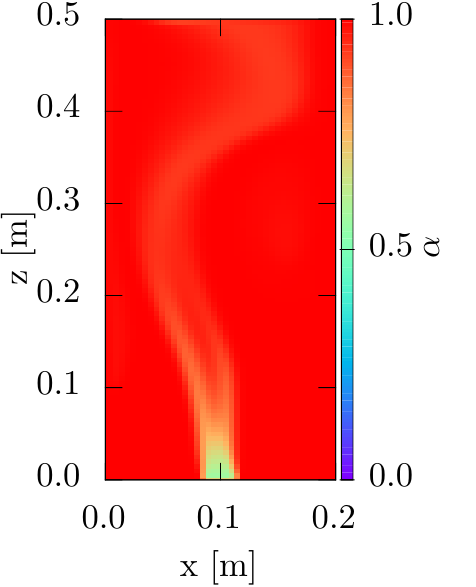}
}\hfill
\subfloat[\label{fig:volFracExcess_noFluc}]{%
   \includegraphics[width=0.25\textwidth]{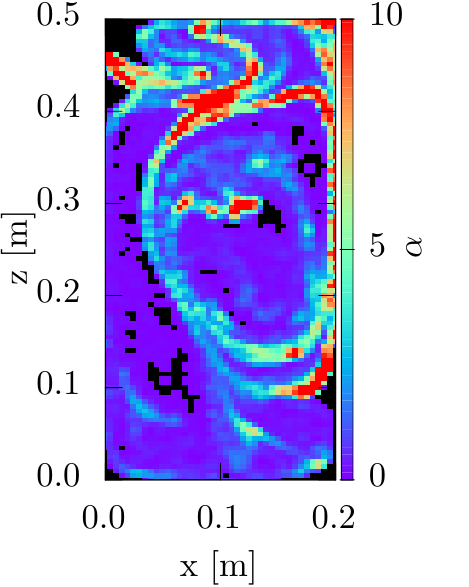}
}\hfill
\subfloat[\label{fig:volFracExcess_fluc}]{%
 \includegraphics[width=0.25\textwidth]{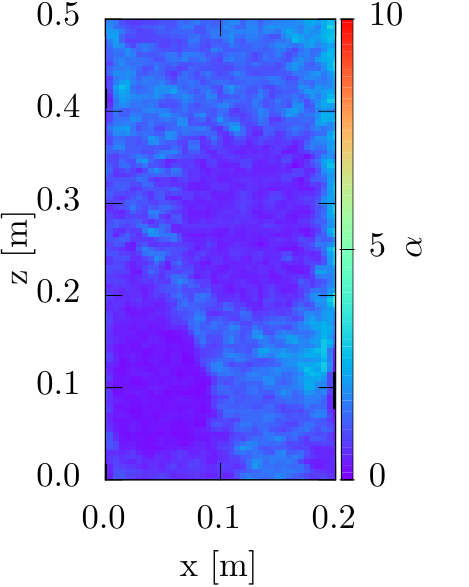}
}
	\caption{Influence of fluctuations on particle velocities and distribution. The first row compares (a) the magnitude of the recurrence velocity with the particle velocity fields (b) without and (c) with fluctuations. Clustering is highlighted in the second row in form of excess of volume fraction fields (e) without and (f) with fluctuations in comparison to the recurrence volume fraction field (d). Completely void regions (black) are present in simulations without fluctuations. Data correspond to a snapshot of the midplane taken at $t=20 \,\second$ with $n_{\text{p}}=3$ and for the cases including fluctuations $D_0=2.5\cdot10^{-3}\meter^2/\second$.} 
	\label{fig:fluctuations}
\end{figure*}

The importance of a proper choice for the fluctuation parameter $D_0$ is demonstrated in Fig.~\ref{fig:fluctuations}. Without or with too small fluctuations, discretization errors cause particles to form clusters and leave regions in-between empty. This happens specifically in vortex regions, where due to their explicit propagation, particles drift outwards and leave the inner region thinly populated. For the case of the bubble column, vortices were formed some distance above the inlet and started to move downwards shortly after. In the course of this process, they pushed highly accumulated particle clusters in front of them towards the bottom where they were fed into the main stream of the bubbles and then gave rise to void regions stemming from their center. Because of the oscillating movement and vortices coming down alternatingly on both sides, too few particles were found on both sides of the bubble stream and too many within it. Fluctuations acted to move some of them into the almost void regions, but if they were too strong, the velocity field was significantly disturbed and the diffusivity artificially increased. Based on the data shown in Fig.~\ref{fig:fluctuations}, we used $n_{\text{p}}=3$ particles per cell and $D_0=2.5\cdot10^{-3}\meter^2/\second$ in our simulations. This choice reduced clustering in form of the volume excess factor from values higher than 10 to not more than 2 and removed (almost) void regions. Note, however, that in order to obtain relaxation towards the recurrence volume fraction field with great accuracy, many more particles per computational cell and rather small time steps would have been needed.

Since conditions with few particles per cell can lead to potential problems with species injection and measurement in such small regions that they might accidentally be empty, a modified insertion/probing mechanism was used. The same amount of tracers per unit time was injected into a somewhat enlarged volume and accumulates until one or more fluid particles passed by to pick it up. For the same reasons, probing can be subject to strong fluctuations.

Both the full CFD as well as all recurrence-based simulations were carried out in parallel on 4 CPUs of an Intel\textsuperscript{\textregistered} Core\texttrademark  i7-2600 processor. Domain decomposition was done equidistantly along the z-axis.

\begin{table}[htb]
	\centering
	\begin{tabular}{llll}
	\toprule
	$\Delta t$ & $\Delta t_{\text{rec}}$ & $\Delta t_{\text{A}}$ & $\Delta t_{\text{B}}$ \\
	\midrule
	$0.0025 \,\second$ & $ 0.1\,\second$ 
		& $ 0.0025\,\second$ & $ 0.025\,\second$ \\
	\bottomrule
	\end{tabular}
	\caption{Comparison of the full CFD, recording, Model A and Model B time steps.}
	\label{tab:timeSteps}
\end{table}

% probes

% probe values

\subsubsection{Results and discussion}
\begin{figure}[htbp]
\centering
\subfloat[\label{fig:species1a}]{%
  \includegraphics[width=0.48\textwidth]{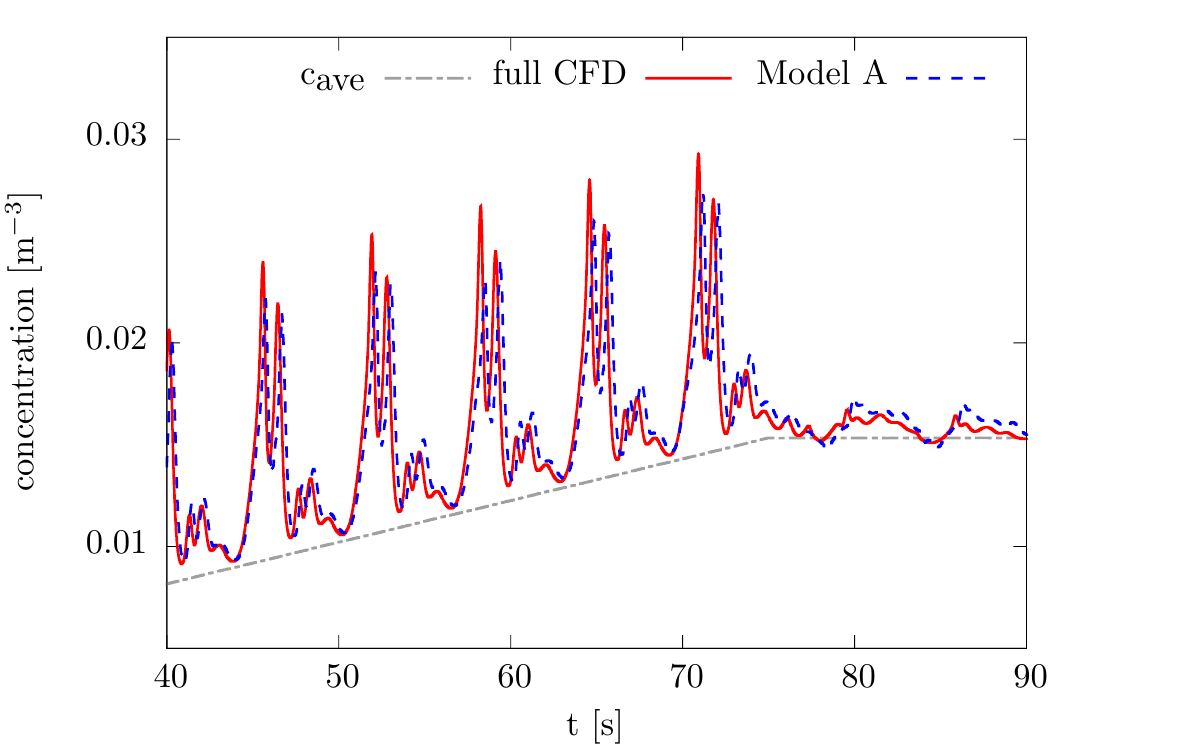}
}\\
\subfloat[\label{fig:species2a}]{%
\includegraphics[width=0.48\textwidth]{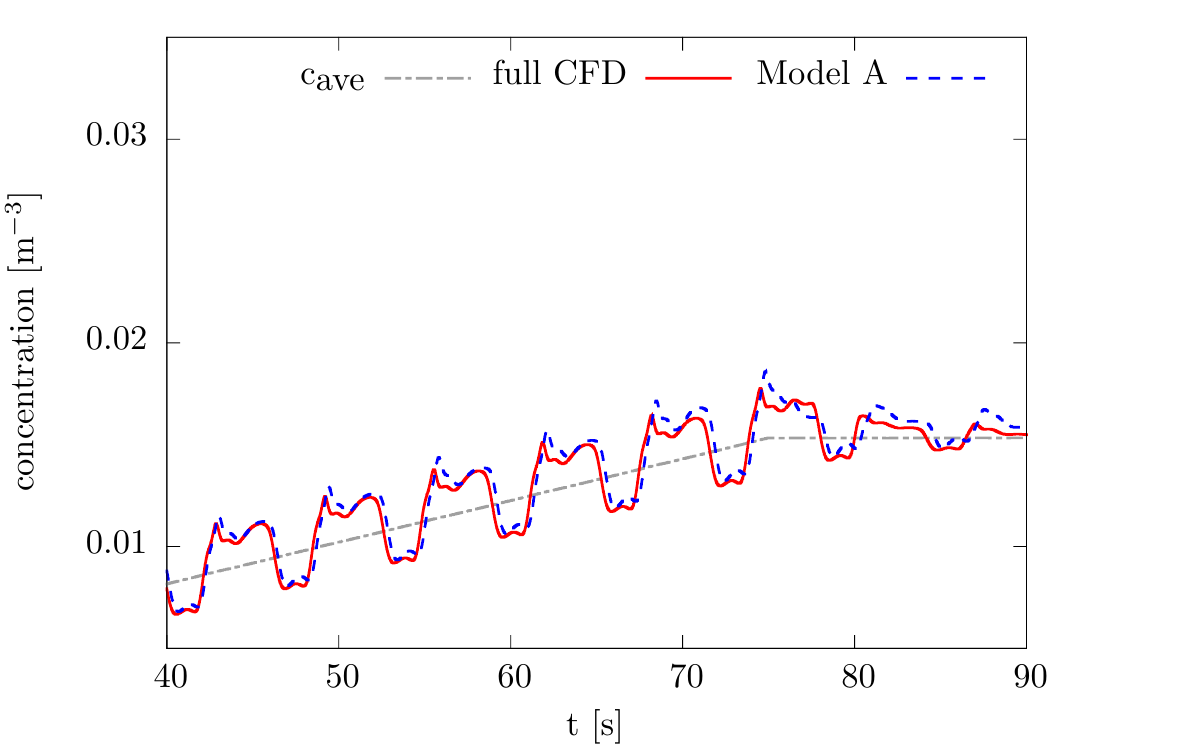}
}\\
\subfloat[\label{fig:species3a}]{%
  \includegraphics[width=0.48\textwidth]{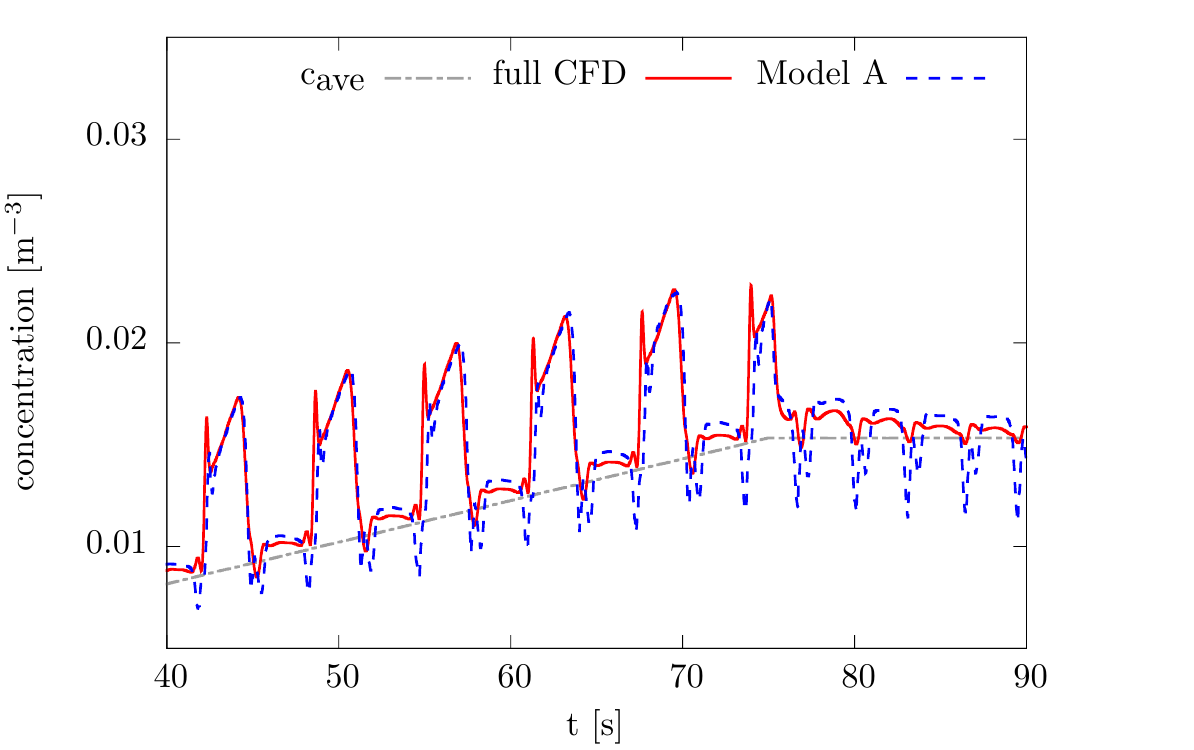}
}
	\caption{\myHighlight{Evolution of species concentrations at locations (a) 1, (b) 2 and (c) 3 defined in Fig.~\ref{fig:pointLocations}. Close inspection reveals that Model A leads to the same two-peak structure as the full CFD calculation. As a guide to the eye, the current, system-wide average of species concentration is displayed, too. For reasons of visibility, we only show the time interval $[40:90]\,\second$, being well beyond recording time and containing the source's switching-off after $ 75\,\second$.}}
	\label{fig:probevaluesA}
\end{figure}

\begin{figure}[htbp]
\centering
\subfloat[\label{fig:species1b}]{%
  \includegraphics[width=0.48\textwidth]{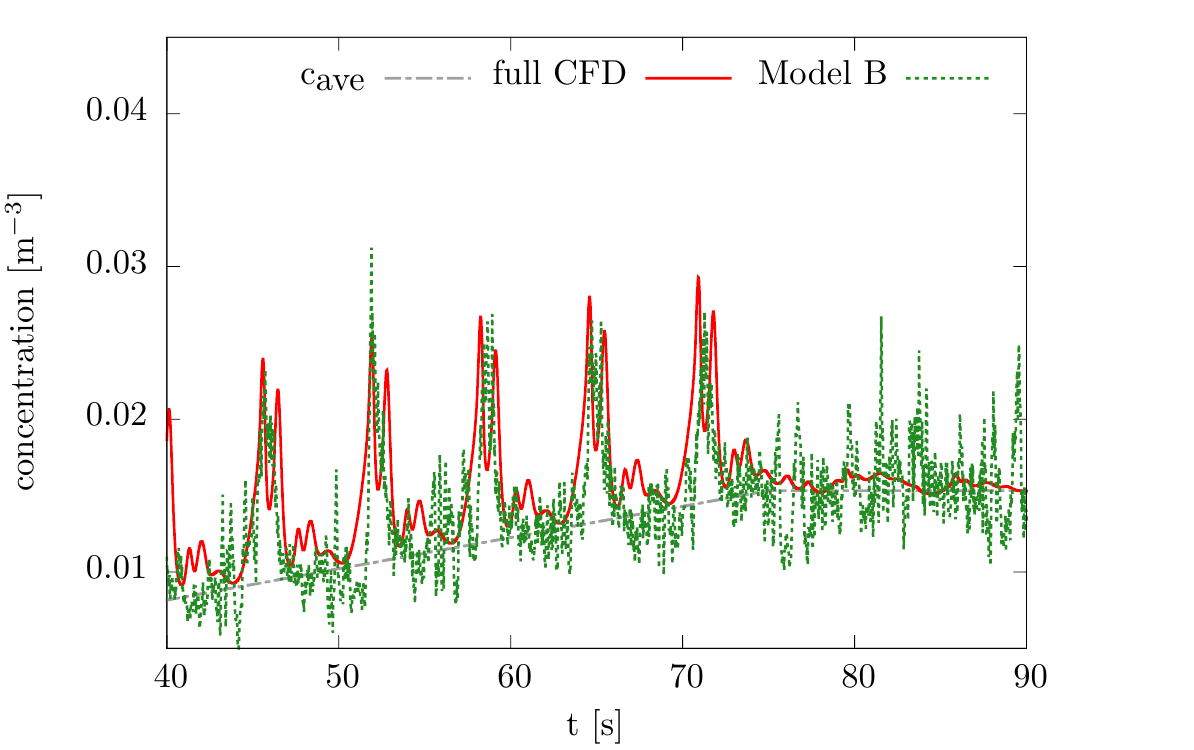}
}\\
\subfloat[\label{fig:species2b}]{%
\includegraphics[width=0.48\textwidth]{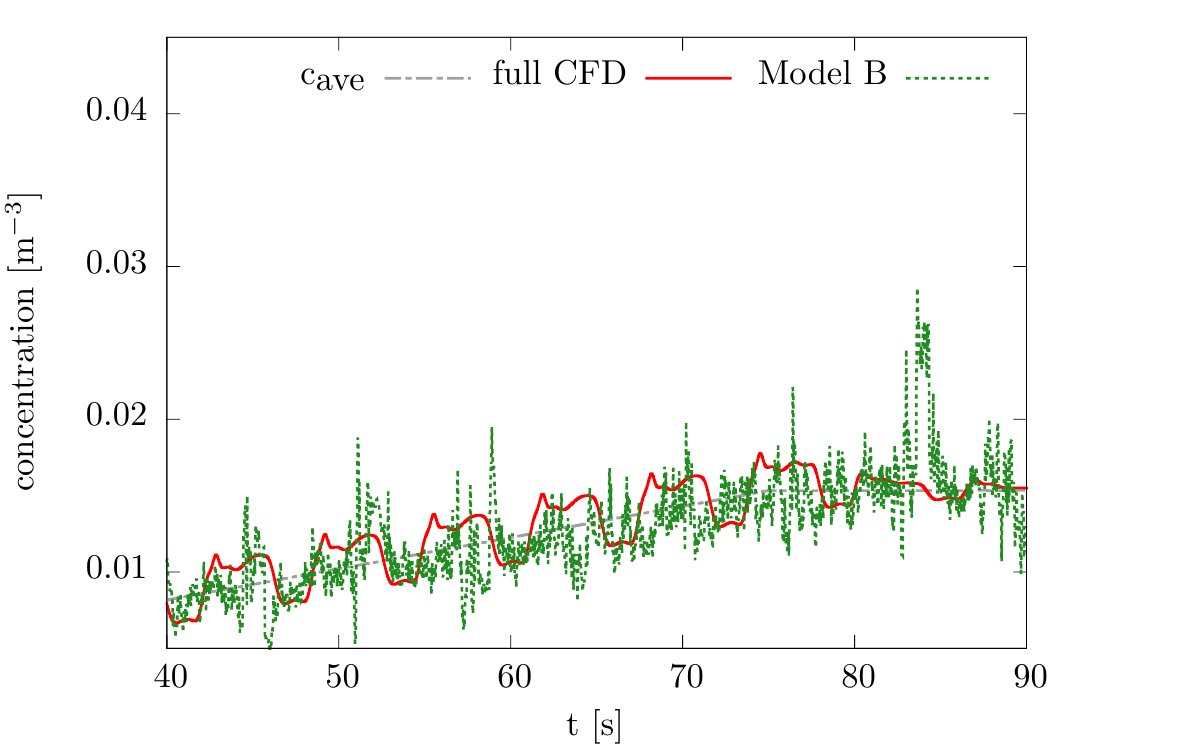}
}\\
\subfloat[\label{fig:species3b}]{%
  \includegraphics[width=0.48\textwidth]{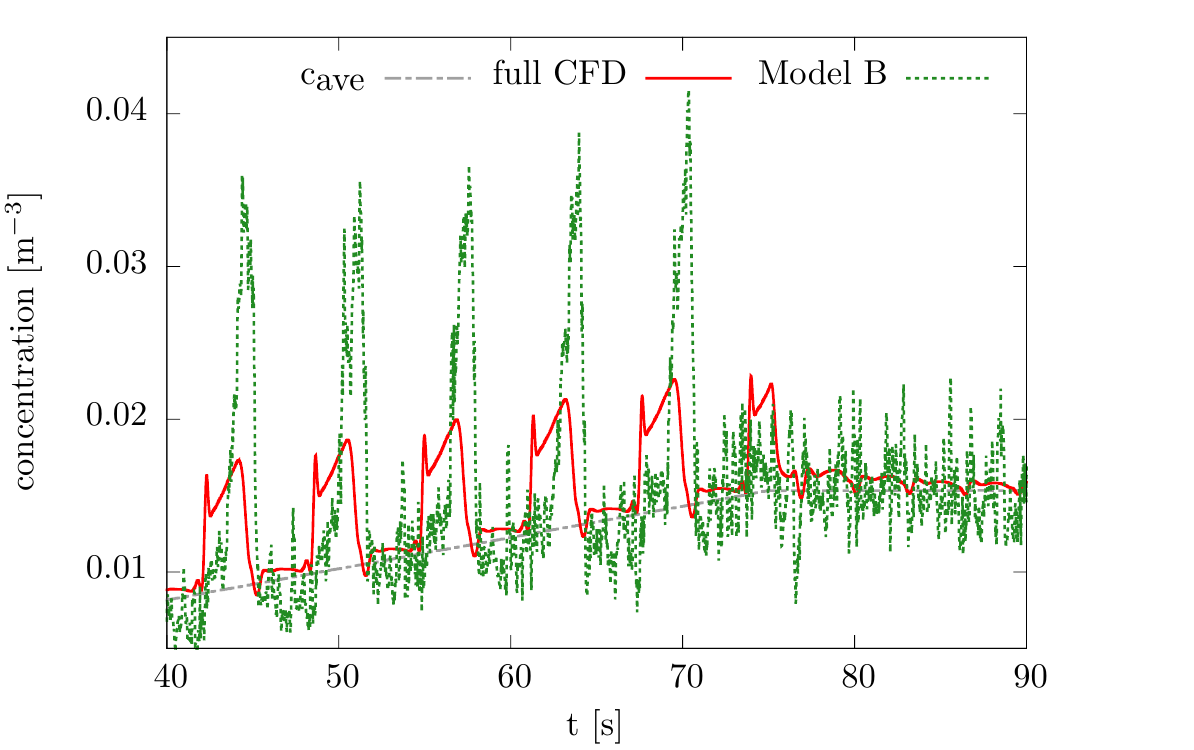}
}
	\caption{\myHighlight{Same as Fig.~\ref{fig:probevaluesA} for Model B. The signal is much noisier and does not resolve details like double-peaks properly but still follows the full calculation roughly.}}
	\label{fig:probevaluesB}
\end{figure}

\myHighlight{Figures~\ref{fig:probevaluesA}  and \ref{fig:probevaluesB}} show the evolution of species concentration at the three probing points specified in Fig.~\ref{fig:pointLocations} together with the average value in the domain. As expected because of the bubble column's back-and-forth motion, we observed two-peak-structures recurring with the bubble stream's period. Peak values increased as long as tracers were injected, after that the signal oscillated around the constant average value. Depending on its location, two different mechanisms could lead to transport to a probing point: either by the main stream or by a vortex moving down beside it. While point 1 was hit directly, point 2 could only be reached by vortex motion. Point 3 first received tracers from the moving-down vortex which were afterwards pushed back from the bubble stream giving rise to its second, more pronounced peak.

\begin{figure}[htbp]
\centering
\includegraphics[width=0.45\textwidth]{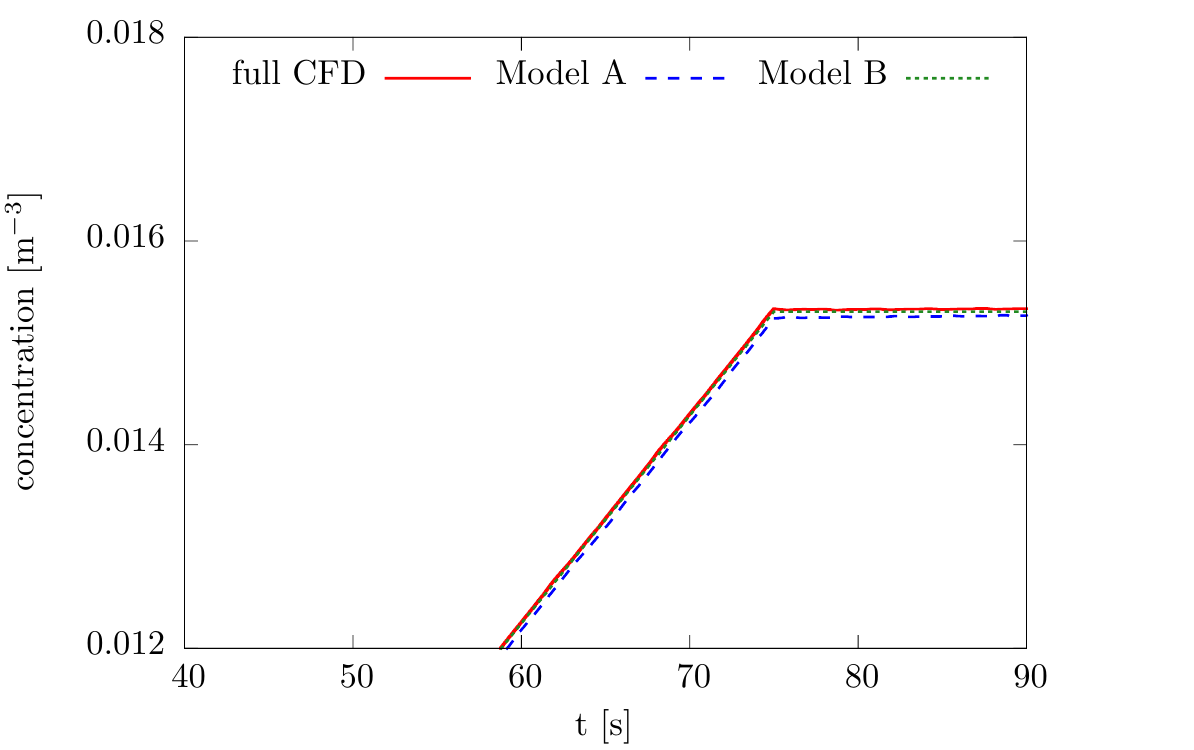}
\caption{\myHighlight{System-wide species average. Both Models A and B agree very well with the full calculation and conserve the injected species globally: as long as the source is turned on, it rises linearly, afterwards it remains constant.}}
 	\label{fig:species_ave}
 \end{figure}

For a highly-recurrent system such as this, Model A captured the transport behavior at all investigated locations very well and showed hardly any recurrence-jump-induced fluctuations. As discussed in section~\ref{sec:recproc}, global conservation demonstrated in Fig.~\ref{fig:species_ave} is a general property of both Models A and B apart from numerical errors. The amount of tracers increased linearly as long as the source was turned on and remained constant afterwards.

Model B was subject to background noise because of particles representing significant fractions of cell volume moving in and out of probing regions. At a first glance, it led to a quite similar peak structure as the full CFD calculation and Model A. However, it \myDelete{missed}\myHighlight{failed} to resolve one of the sub-peaks. At point 1, just the first one was visible. Here, the bubble column was moving leftwards bringing fluid parcels with its main stream into a previously thinly populated region. As long as there was no excess of particles, no fluctuations disturbed transport. However, on moving back rightwards, too many parcels had accumulated and strong fluctuations smeared out the second peak. A similar behavior could be observed at point 3. First, a dense corona was shoved through the point from a downwards moving vortex. The present fluctuations prevented a visible concentration peak. After that, point 3 lay within the vortex and tracers being pushed back by the main stream were not deflected. As a matter of fact, the bubble column collected \myDelete{a too high}\myHighlight{an excessive} number of them leading to an increased peak value. A similar mechanism was responsible for the shape of the signal at point 2.

\begin{figure}[htbp]
\centering
\subfloat[\label{fig:species_D0}]{%
  \includegraphics[width=0.45\textwidth]{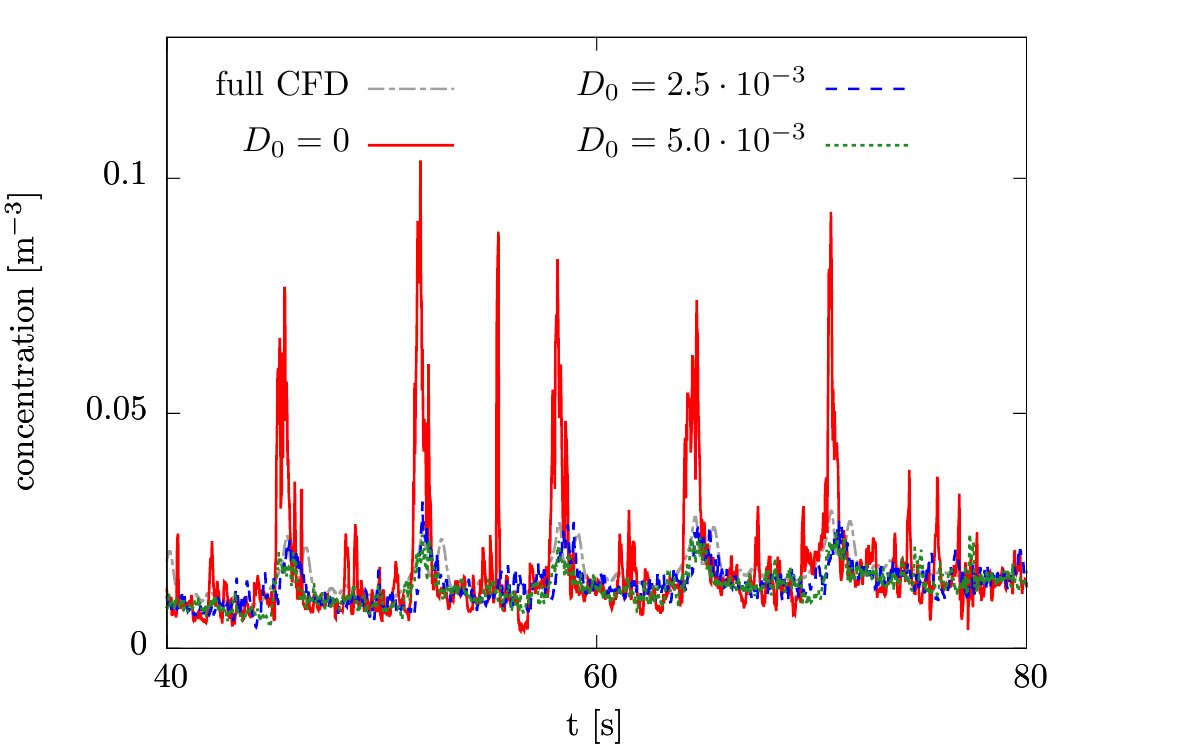}
}\hfill
\subfloat[\label{fig:species_np}]{%
\includegraphics[width=0.45\textwidth]{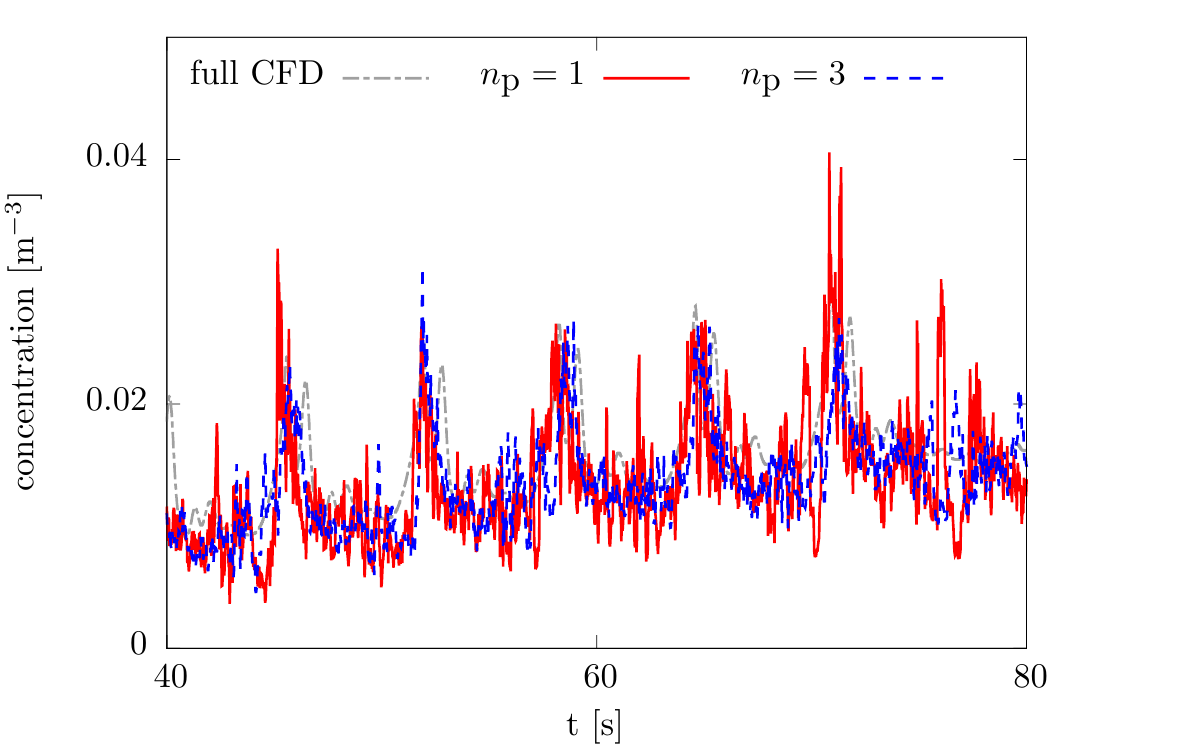}
}
	\caption{Influence of simulation parameters in Model B. (a) highlights the importance of parcel fluctuations to prevent artificial spikes in transport signals ($n_\text{p}=3$ for all curves). It is shown in (b) that the number of particles per computational cell slightly influences the amount of background noise but does not change results qualitatively ($D_0=2.5\cdot 10^{-3}$ for all curves). }
	\label{fig:paraB}
\end{figure}

These considerations show that while fluctuations help to avoid empty regions preventing transport, they inhibit the investigation of spatially highly resolved phenomena. Too strong fluctuations smear out relevant features whereas too weak ones might leave the same unsampled and impose additional noise, which is also emphasized by Fig.~\ref{fig:species_D0}. Since this dilemma cannot be \myDelete{cured with}\myHighlight{resolved by} using more particles to represent a cell and \myHighlight{since} much smaller time steps would quickly render the method impractical, it turns out that Model B is most suitable for systems without strongly localized features. Then, however, it is permissible to choose a rather coarse discretization. It is shown in Fig.~\ref{fig:species_np} that apart from a somewhat increased background noise, one particle per cell led to qualitatively very similar results as three, but significantly reduced the run time.

\begin{figure}[htbp]
\centering
  \includegraphics[width=0.45\textwidth]{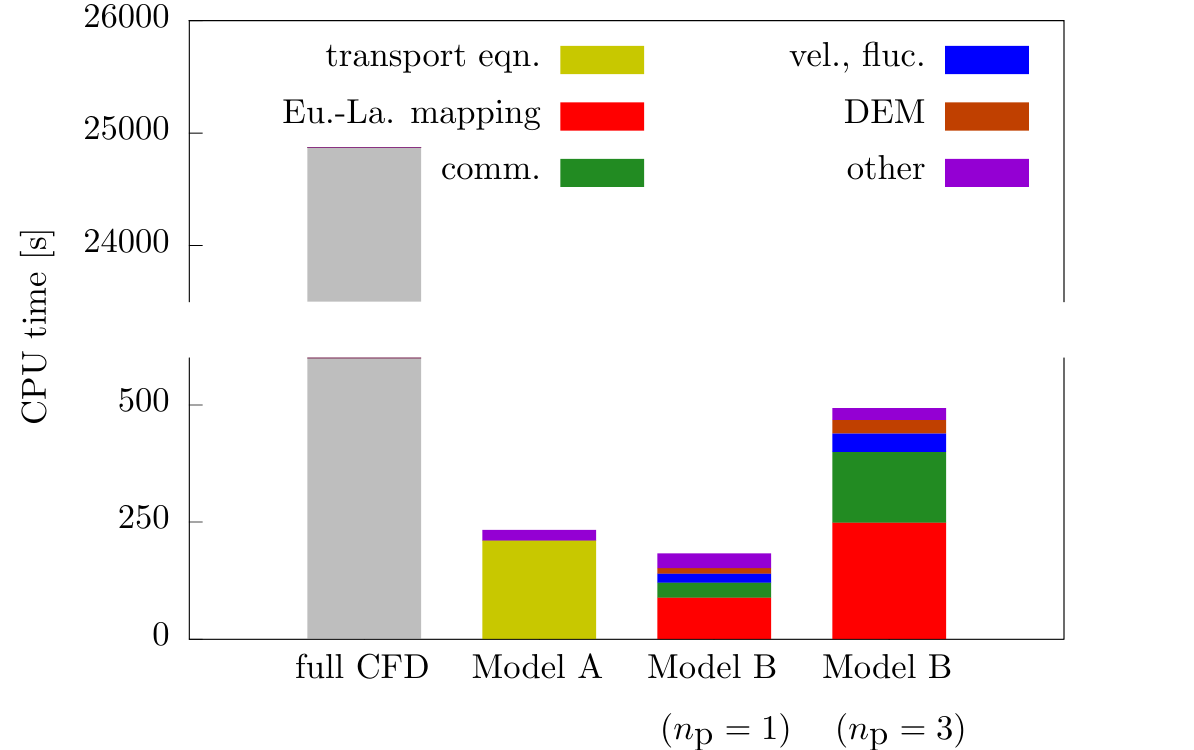}
	\caption{Overview of the various contributions to CPU time measured after the simulations' initialization phase. The most important sources are the solution of the transport equation for Model A and the mapping of Lagrangian particle information on Eulerian fields, communication between the CFD and DEM solvers, calculation of particle velocities and fluctuations as well as the DEM solver's particle propagation for Model B. Each simulation was carried out on 4 CPUs of an Intel\textsuperscript{\textregistered} Core\texttrademark  i7-2600 processor.}
	\label{fig:exTime}
\end{figure}

A comparison of simulation times of Models A and B with the full calculation measured after the programs' initialization phase as well as an analysis of their main sources are displayed in Fig.~\ref{fig:exTime}. Since wall-clock and CPU time differed by not more than a second for all cases, we chose to show only the latter.
As expected, recurrence-based methods outperformed a corresponding conventional simulation by \myDelete{magnitudes}\myHighlight{orders of magnitude}. For Model A, almost all CPU time was spent with solving the scalar transport equation. Other contributions, stemming e.g.\ from monitoring the amount of species, were negligible. For Model B, a larger variety of sub tasks was present. Most time was spent with mapping Lagrangian particle information on Eulerian grids to obtain volume fraction fields required to calculate velocity fluctuations. The second-largest term was due to communication between the CFD solver handling field data and the DEM solver responsible for particle propagation. Together with the time required to evaluate velocities and fluctuations and by the DEM solver to update particle positions, they constituted the main part of Model B's simulation time. Clearly, all of these terms increased with particle number. The presence of a possibly large data base of recurrence fields causes an insignificantly extended initialization phase during which they are loaded into memory. Since this happens only once for each simulation run, it is not shown in Fig.~\ref{fig:exTime}.

In the present case, Model A and B led to approximately the same simulation time if each cell was represented by one particle. We stress, however, that both the mapping procedure and the communication between the CFD and DEM solver are suitable for future optimization to further improve Model B's performance.

% probe values

% % probe values
% \begin{figure}[htbp]
% \centering
% \subfloat[\label{fig:species1}]{%
%   \includegraphics[width=0.48\textwidth]{images/species1_c.pdf}
% }\hfill
% \subfloat[\label{fig:species2}]{%
% \includegraphics[width=0.48\textwidth]{images/species2_c.pdf}
% }\\
% \subfloat[\label{fig:species3}]{%
%   \includegraphics[width=0.48\textwidth]{images/species3_c.pdf}
% }\hfill
% \subfloat[\label{fig:species_ave}]{%
% \includegraphics[width=0.48\textwidth]{images/species_ave.pdf}
% }
% 	\caption{Evolution of species concentrations at locations (a) 1, (b) 2 and (c) 3 defined in Fig.~\ref{fig:pointLocations}. Close inspection reveals that Model A leads to the same two-peak structure as the full CFD calculation while Model B predicts one larger peak. (d) shows the global conservation of the injected species.}
% 	\label{fig:probevalues}
% \end{figure}

% B parameters

% profile

\subsection{Steelmaking BOF Converter Process}
\label{sec:bof}
\begin{figure*}[htbp]
\centering
  \includegraphics[width=0.85\textwidth]{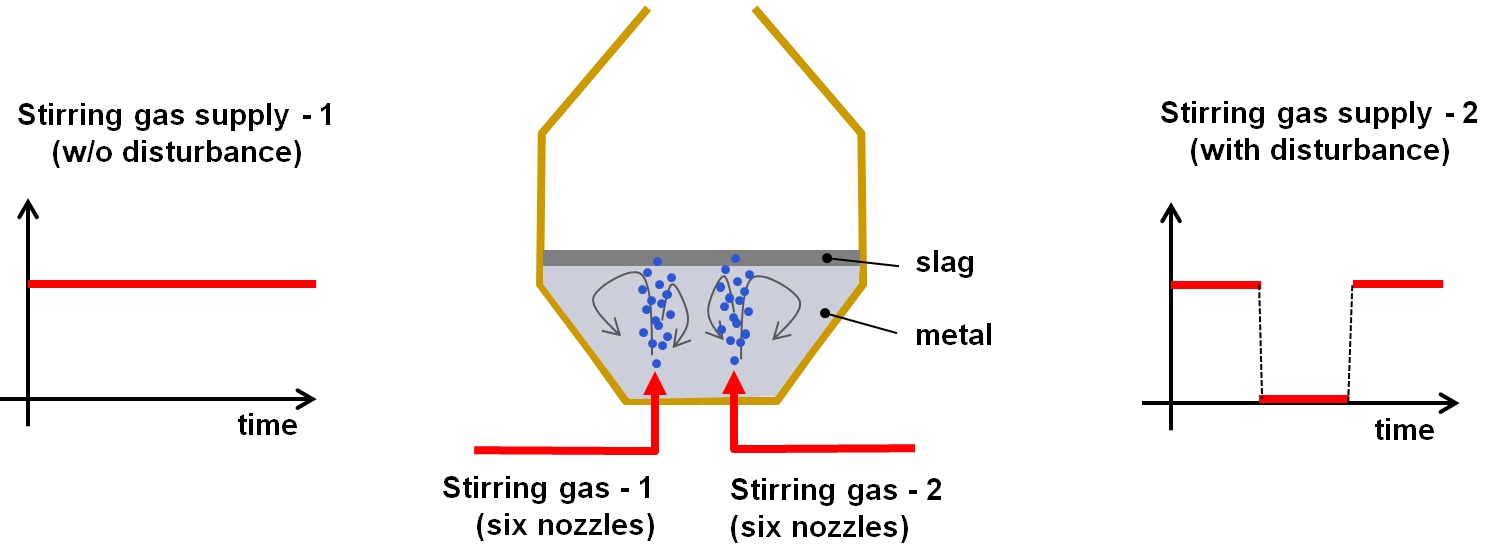}
	\caption{Sketch of the BOF converter flow configuration and operation. As assumed in this application example, the second gas supply line is shut down between the second and third minute of operation.}
	\label{fig:bofconv}
\end{figure*}

In a second application we considered recurrence CFD simulations of steel decarbonisation during post-stirring in a Basic Oxygen Furnace (BOF). In this common steelmaking process step, liquid metal is covered by liquid slag while mixing is achieved by bubble plumes which are introduced through porous plugs at the bottom of the vessel (see Fig.~\ref{fig:bofconv}). In addition to the homogenization of the steel melt, this flow agitation triggers a transport-controlled heterogeneous reaction at the metal-slag interface. 

In a first step we addressed this multiphase flow configuration by conventional CFD simulations, modelling the two stratified fluids (metal and slag) by a Volume of Fluid (VoF) method, the bubble plumes by tracking representative Lagrangian parcels of bubbles and fluid turbulence by a coarse grained Large Eddy Simulation (LES).
More details on this modelling concept and on cold-water validation experiments with the aim to study converter oscillations can be found in \cite{Pirker2012}.

On top of this multiphase flow, we considered decarbonisation at the metal-slag interface where carbon reacts with iron-oxide to iron and carbon-monoxide,
\begin{equation}
 [C]+\langle FeO \rangle \rightarrow [Fe] + \{CO\}.
\end{equation}
Brackets $[]$, $\langle \rangle$ and $\{ \}$ indicate the state of dissolution, i.e.\ metal dissolved, slag dissolved and gaseous. The reaction kinetics is subject to a ``mass acting'' first-order law,
\begin{equation}
 \dot{N}=\kappa A_{\text{r}}\Big([C]\langle FeO\rangle - [C]\langle FeO\rangle\Big|^{(\text{eq})} \Big).
\end{equation}
The molar reaction rate $\dot{N}$ is controlled by three terms. The most important last term describes the mass acting degree of non-equilibrium, i.e.\ the difference between the current concentrations of the reactants and their equilibrium concentrations. The second term $A_{\text{r}}$ describes the available reaction area between metal and slag and the mass transfer coefficient $\kappa$ models sub-grid transport processes.

We further assumed that the available reaction area is significantly influenced by local emulsification such that
\begin{equation}
 A_{\text{r}} =
 \begin{cases}
 A_{\text{r}}^{(\text{strat})}=4\alpha_{\text{m}}\alpha_{\text{s}}V_c^{2/3} &\mbox{if } \dot{\gamma} < \dot{\gamma}_{\text{emul}} \\
A_{\text{r}}^{(\text{emul})}=6\alpha_{\text{m}}\alpha_{\text{s}}V_c /
\big((\alpha_{\text{m}}+\alpha_{\text{s}})d_{\text{b}}\big) & \mbox{if } \dot{\gamma} \geq \dot{\gamma}_{\text{emul}}
\end{cases}
\end{equation}
with $\alpha_{\text{m}}$ and $\alpha_{\text{s}}$ being the phase volume fractions of liquid metal and slag, $V_c$ is the volume of the local grid cell and $d_{\text{b}}$ is a typical emulsion bubble diameter for which we chose $d_{\text{b}}=5\,\milli\meter$. Phase interaction was controlled by the local shear rate $\dot{\gamma}$ with the limiting shear rate set to $\dot{\gamma}_{\text{emul}}=10^{-2}\, \second^{-1}$.

At this point we accepted this modelling concept as a reasonable description of steel decarbonisation in a steelmaking converter. We further boldly simplified the reaction rate to
\begin{equation}
 \dot{N}=\kappa A_{\text{r}}\Big([C] - [C]\Big|^{(\text{eq})} \Big)
\end{equation}
with $[C]^{(\text{eq})}=3.0\, \text{wt}\%$ for the sake of simplicity.

Full CFD simulations of steel decarbonisation during a process period of three minutes were performed on a $400k$ purely hexahedral grid. Carbon concentration was initialized with $[C]^{(0)}=6.0\, \text{wt}\%$ throughout the melt. We assumed that in the second minute, half of the stirrers are shut down due to a disturbance in the second gas supply line, before they are switched on again in the third minute (see Fig.~\ref{fig:bofconv}). Due to the highly dynamic flow pattern, very small time-steps in the order of $\Delta t = 10^{-3}\, \second$ had to be applied, resulting in an overall simulation time of approximately two weeks for a three minutes process.
These full CFD simulations serve as a validation base for our recurrence CFD approach. 

\subsubsection{Set-up of Recurrence CFD}
We obtained recurrence plots for both process states (i.e.\ all stirrers in operation and only half of stirrers in operation) from corresponding full CFD simulations.
The recording time-step was linked to the dynamics of characteristic flow features (e.g.\ bubble clusters passing through the melt) by defining ${\Delta t_{\text{rec}}=0.05\,\second}$. Data were recorded for a period of $\tau_{\text{rec}}=5\,\second$, resulting in $N_{\text{rec}}=100$ data-frames for each process state. In each data-frame, the vector
\begin{equation}
 \textbf{x}_{\text{rec}}=\Big(\alpha_{\text{m}},\, \alpha_{\text{s}},\, \textbf{u},\, A_{\text{r}}\Big)
\end{equation}
was stored. Note, that in addition to the phase volume fractions and the velocity field we also stored the locally available reaction area $A_{\text{r}}$ because it was needed for the calculation of the instantaneous reaction rate.

\begin{figure}[htbp]
\centering
\subfloat[\label{fig:recMat2A}]{%
  \includegraphics[height=0.3\textwidth]{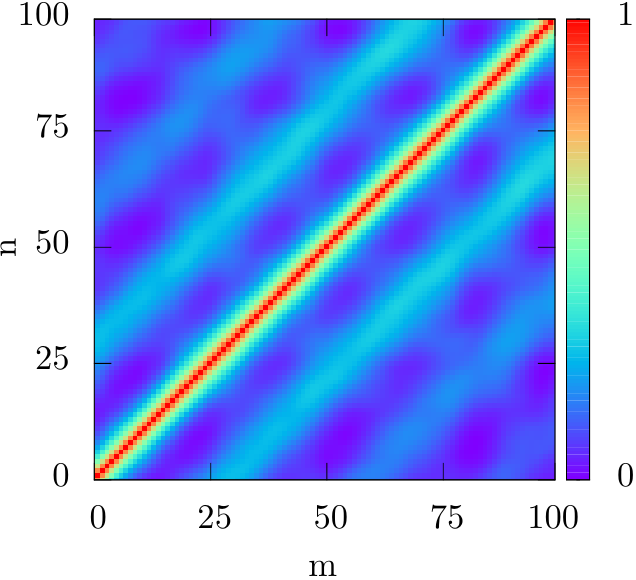}
}\hfill
\subfloat[\label{fig:recMat2B}]{%
\includegraphics[height=0.3\textwidth]{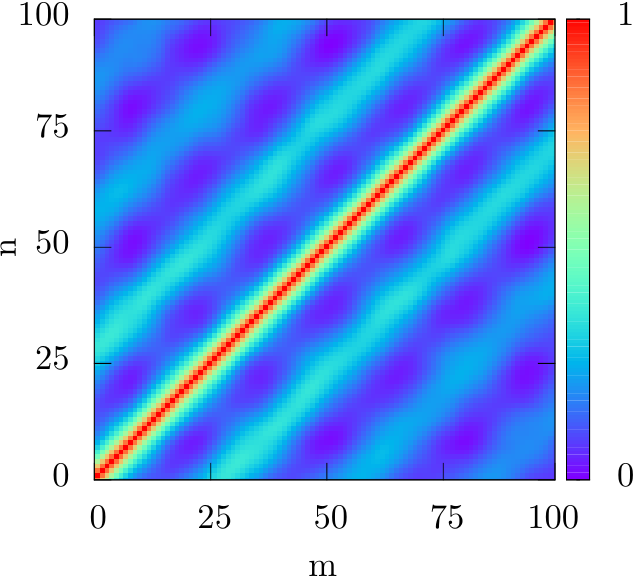}
}
\caption{Recurrence plots of multiphase simulations of the BOF steelmaking process with (a) undisturbed and (b) disturbed gas supply.}
\end{figure}

The recurrence plots in Figs.~\ref{fig:recMat2A} and \ref{fig:recMat2B} indicate that the process considered in this application is less deterministic than the oscillating bubble column in section~\ref{sec:obc}. In contrast to the recurrence plots in Figs.~\ref{fig:recMat_alpha} and \ref{fig:recMat_phi}, recurring flow features are not that regular, which can be related to less pronounced periodic flow behavior.

Based on these recurrence statistics, we composed generic flows from recurrence processes as described in section~\ref{sec:recproc}. \myDelete{Finally, we applied Model B in order to picture the decarbonisation process. We opted for the Lagrangian description because the less regular structure of the recurrence plots indicated that stitching individual flow sequences would not be as smooth as in the case of the bubble column. In Model A this might have led to an accumulation of local mass-imbalances at the stitching instances while such imbalances were avoided by Model B due to its substantial nature.}\myHighlight{Finally, we applied recurrence CFD in order to picture the decarbonisation process. In a first implementation of Model A, we experienced bad numerical convergence especially when switching the recurrence process (accounting for changing operation conditions). Most probably, these numerical problems can be associated with local mass imbalances which are caused by stitching individual flow sequences. In Model A, these imbalances deteriorate the convergence behavior of the species transport equation. Model B, in turn, avoids such imbalances due to its substantial nature. Consequently, we opted for a Model B implementation of recurrence CFD in this case.}

\subsubsection{Results of Recurrence CFD}
\begin{figure*}[htbp]
\centering
  \includegraphics[height=0.38\textwidth]{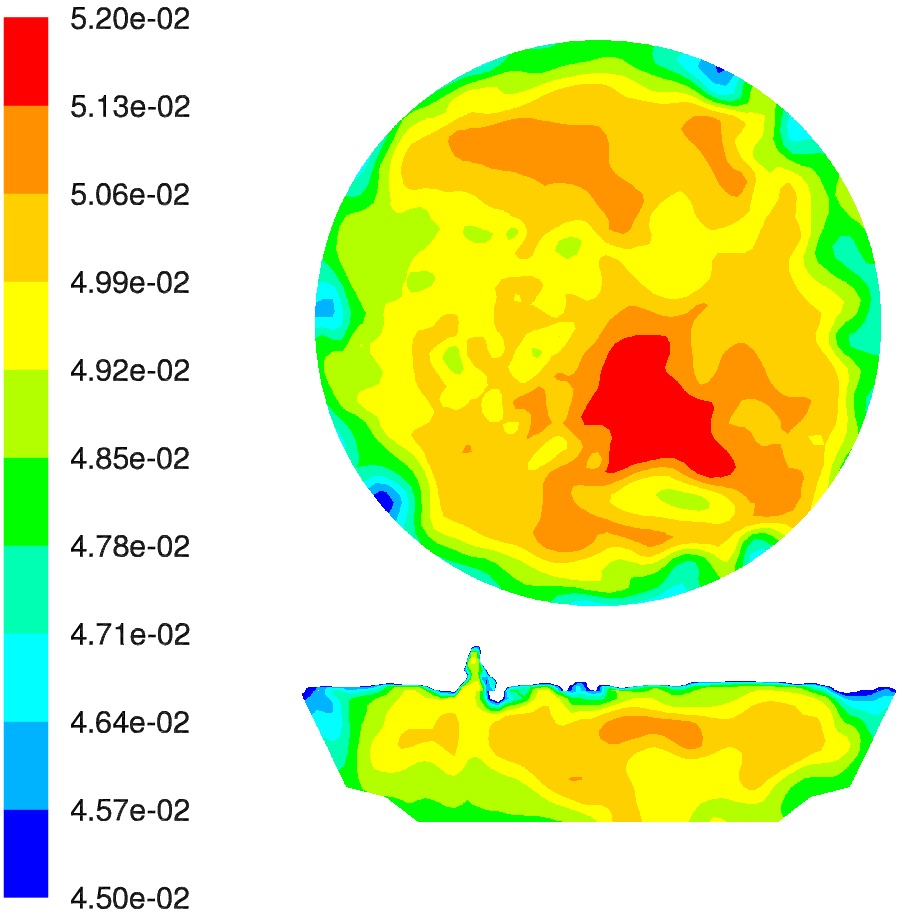}
\qquad
\includegraphics[height=0.38\textwidth]{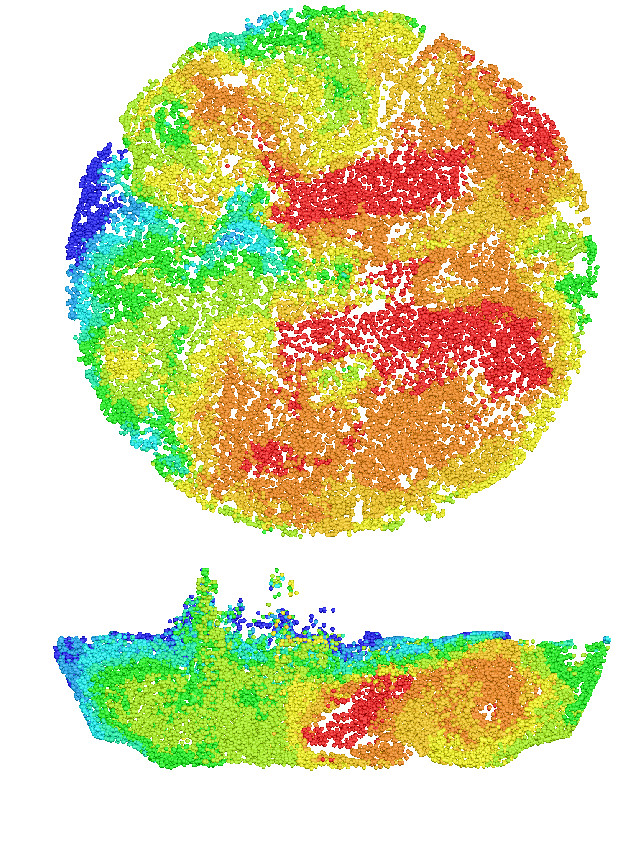}
\caption{\myDelete{Carbon concentration in a (top) horizontal and (bottom) vertical observation plane as result of (left) full CFD simulations and (right) recurrence CFD simulations. Snapshots were taken at two minutes process time (end of disturbance, see Fig.~\ref{fig:bofconv}) with only stirrers on the left hand side in operation.}\myHighlight{Carbon concentration in a (top) horizontal and (bottom) vertical observation plane as result of (left) full CFD simulations and (right) recurrence CFD simulations (250k particles) at two minutes process time (end of disturbance, see Fig.~\ref{fig:bofconv}). Only stirrers on the left-hand side are in operation.}}
\label{fig:carbonconc}
\end{figure*}

In Fig.~\ref{fig:carbonconc} the process of steel decarbonisation is illustrated with the carbon concentration in a horizontal and a vertical observation plane. While in case of full CFD, Eulerian concentration fields are shown, Lagrangian fluid particle concentrations are plotted for recurrence CFD. Note that in the latter case, fluid particles are depicted within an observation layer rather than at a distinct plane. The snap-shots presented in this figure were taken at the end of the disturbance period, with half of the stirrers still out of operation. While there were significant differences in the shapes of the concentration minima and maxima, both simulation approaches – full CFD and recurrence CFD – agreed in predicting a globally asymmetric concentration pattern. It can be clearly seen that the shutdown of the (in this view) right stirrers led to a globally asymmetric distribution of dissolved carbon, with enriched carbon concentration in the unstirred vessel region. 

Due to the coarse particle resolution used, this run of recurrence CFD exhibited a maldistribution of fluid particles, which can be detected by the white, particle-free regions within the plots on the right-hand side of Fig.~\ref{fig:carbonconc}. It should be noted that in case of higher particle resolutions, the homogeneity of particle distribution would improve significantly.

\begin{figure}[htbp]
\centering
  \includegraphics[height=0.35\textwidth]{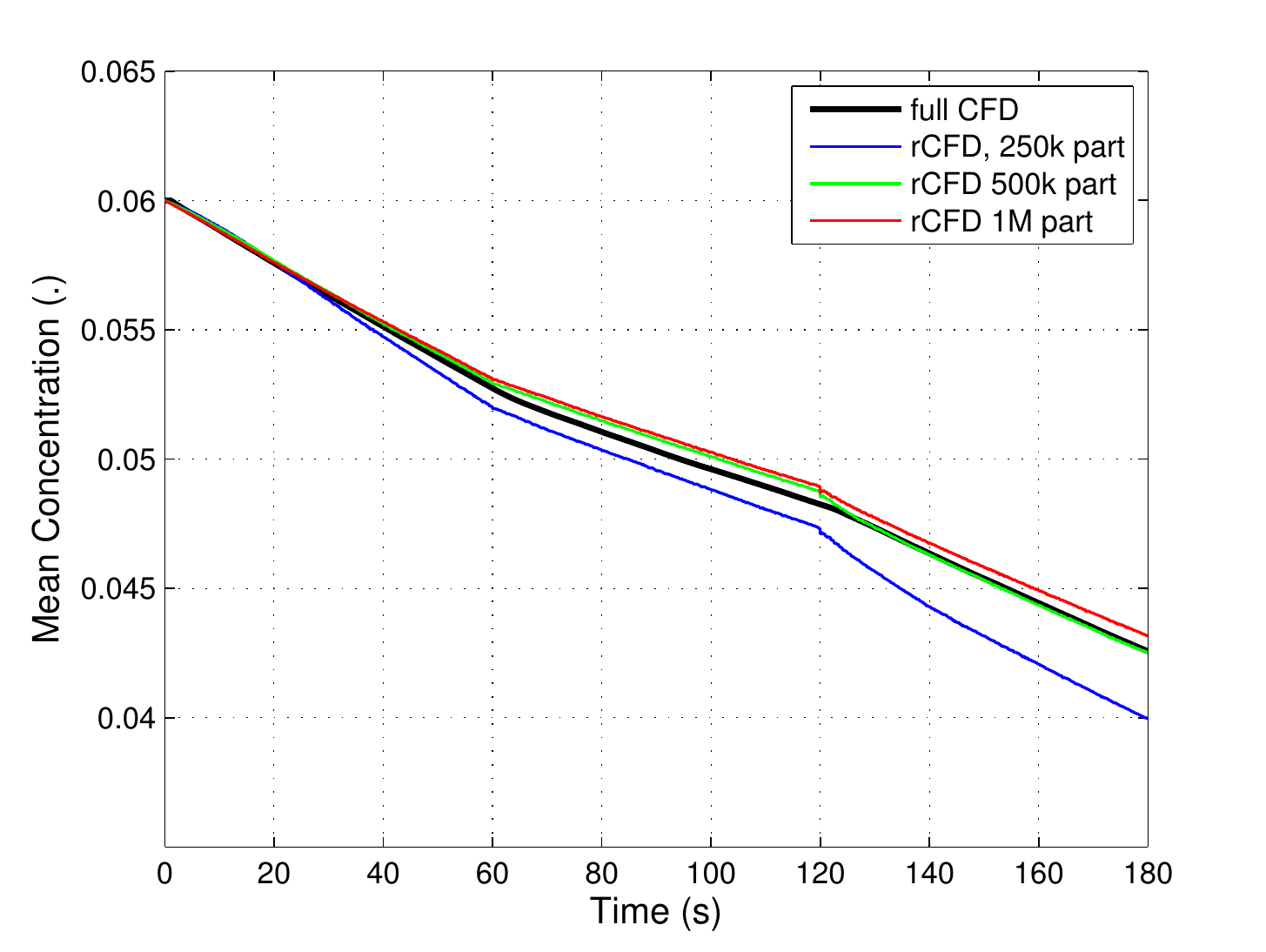}
\caption{Decay of mean carbon concentration with time resulting from full (black line) and recurrence CFD with different resolutions (blue, green and red lines).}
\label{fig:concplot1}
\end{figure}

\begin{figure}[htbp]
\centering
  \includegraphics[height=0.35\textwidth]{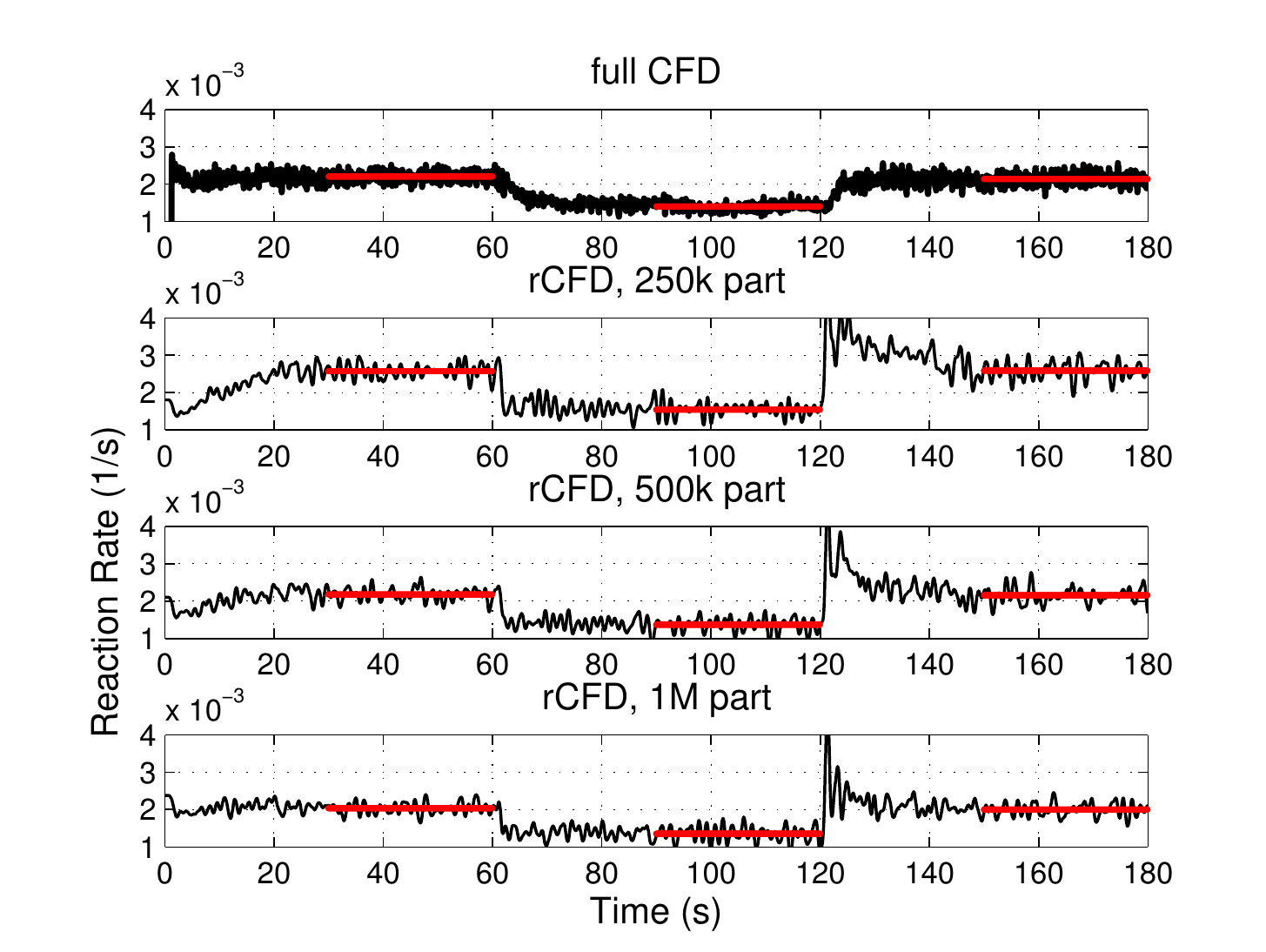}
\caption{\myDelete{Instantaneous decarbonisation rates over process time as a result of full (top) and recurrence CFD with different resolutions (three bottom lines). Note the temporal overprediction of recurrence CFD at the beginning of the last process phase.}\myHighlight{Instantaneous decarbonisation rates over process time as a result of full (top) and recurrence CFD with different resolutions (three bottom lines). Note the temporal overprediction of recurrence CFD at the beginning of the last process phase. Maximum overshoots not visible in this figure reach $5.5$, $5$, $5\cdot10^{-3}$ for 250k, 500k and 1M particle resolution, respectively.}}
\label{fig:concplot2}
\end{figure}

In addition to this qualitative observation, Figs.~\ref{fig:concplot1} and \ref{fig:concplot2} deliver quantitative monitors of mean carbon concentration decay and instantaneous reaction rate, respectively. Both results indicate that the predictive capability of recurrence CFD can be improved by finer resolution, i.e.\ by inserting more Lagrangian fluid particles. While a coarse discretization with $250k$ particles significantly overpredicted decarbonisation, finer discretization with $500k$ or $1M$ particles led to better agreement.

Overall, these concentration curves prove that recurrence CFD is actually able to  correctly predict decarbonisation rates for quasi-stationary process states. In Fig.~\ref{fig:concplot2} this agreement is highlighted by adding a solid average line to the monitors of instantaneous reaction rates for each second half of a minute. While changing the process state led to temporary disagreement, recurrence CFD quickly relaxed towards the results of full CFD once the new process state was established. We could further observe that this relaxation period became smaller with higher fluid particle resolution.

\begin{figure}[htbp]
\centering
\subfloat[]{%
  \includegraphics[height=0.35\textwidth]{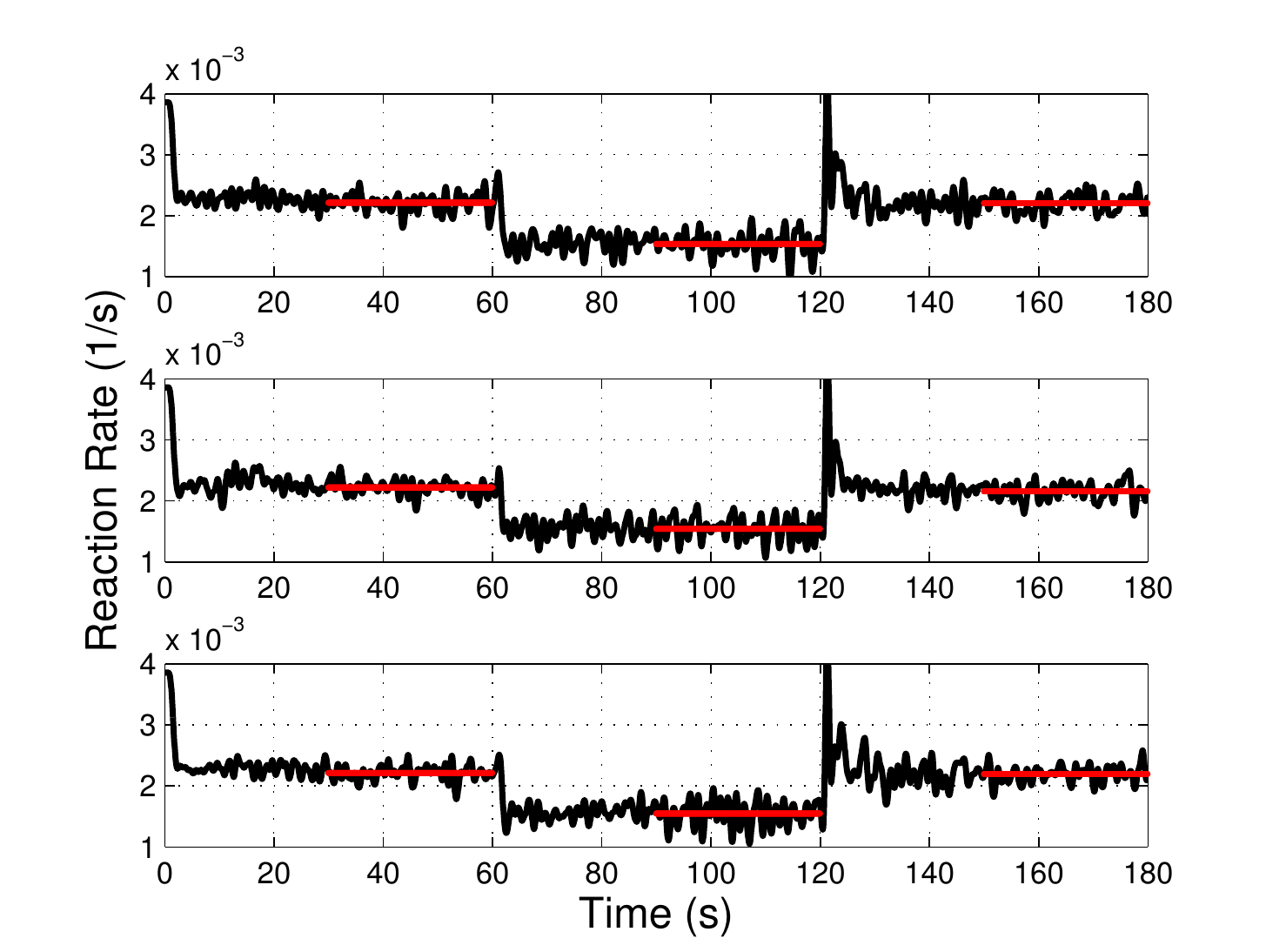}
}\\
\subfloat[]{%
\includegraphics[height=0.35\textwidth]{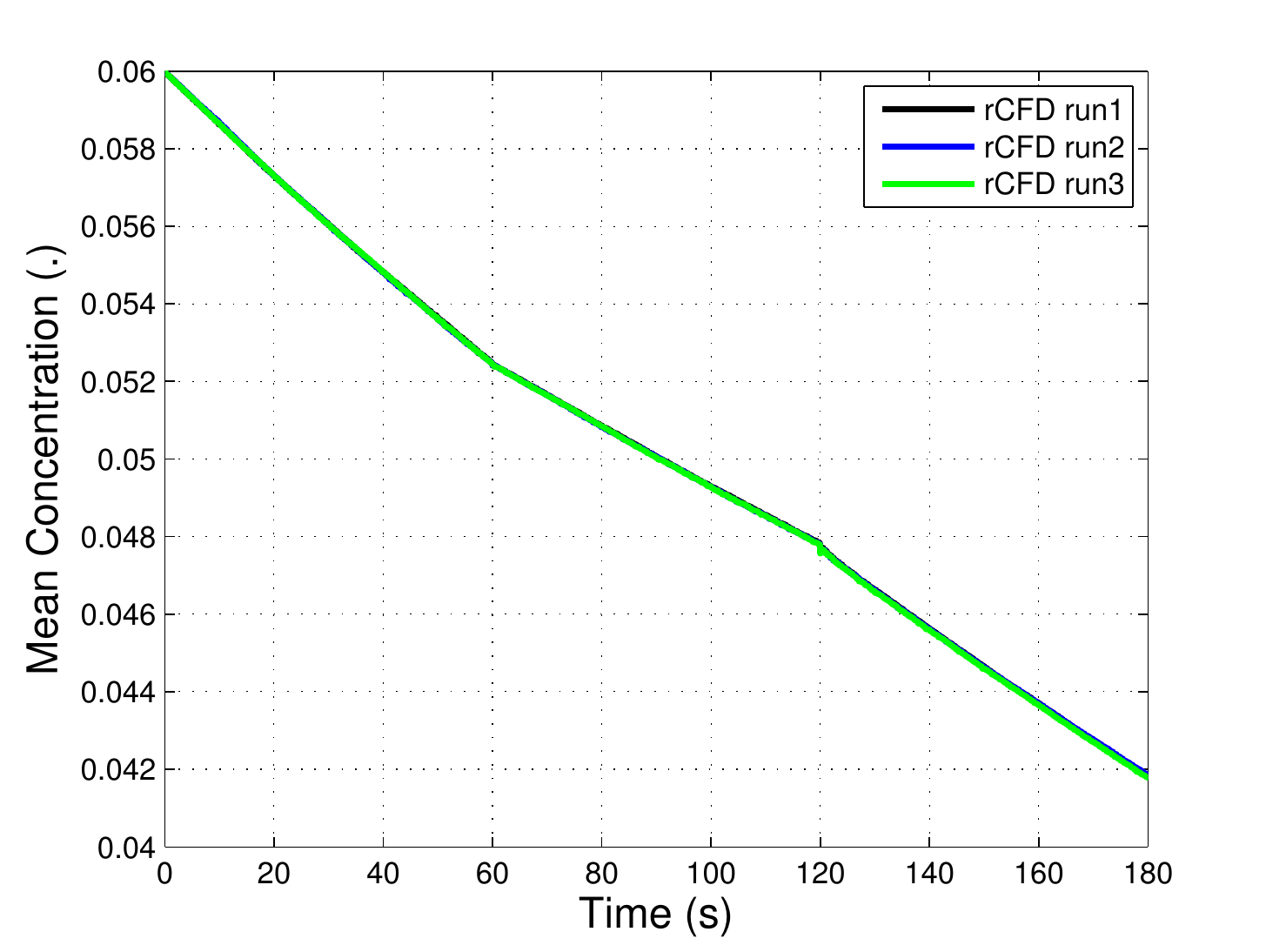}
}
\caption{Repeatability of recurrence CFD simulations. Unique monitors of instantaneous reaction rates (top) result in identical global carbon concentration decay (bottom).}
\label{fig:concplot3}
\end{figure}

With Fig.~\ref{fig:concplot3} we want to emphasize the random nature of recurrence CFD. Here, reaction rates were obtained by re-running recurrence CFD three times with the very same settings. Due to the random nature of the recurrence process, each run is unique, which led to visible deviations in the instantaneous reaction rates (which can be most easily detected at the beginning of the third process period). Interestingly, the global carbon concentration decay depicted in the lower part of Fig.~\ref{fig:concplot3} was completely independent of those instantaneous deviations. Obviously, the global process of decarbonisation was not sensitive to the actual sequencing of characteristic flow pattern.

\begin{figure*}[htbp]
\centering
  \includegraphics[width=0.75\textwidth]{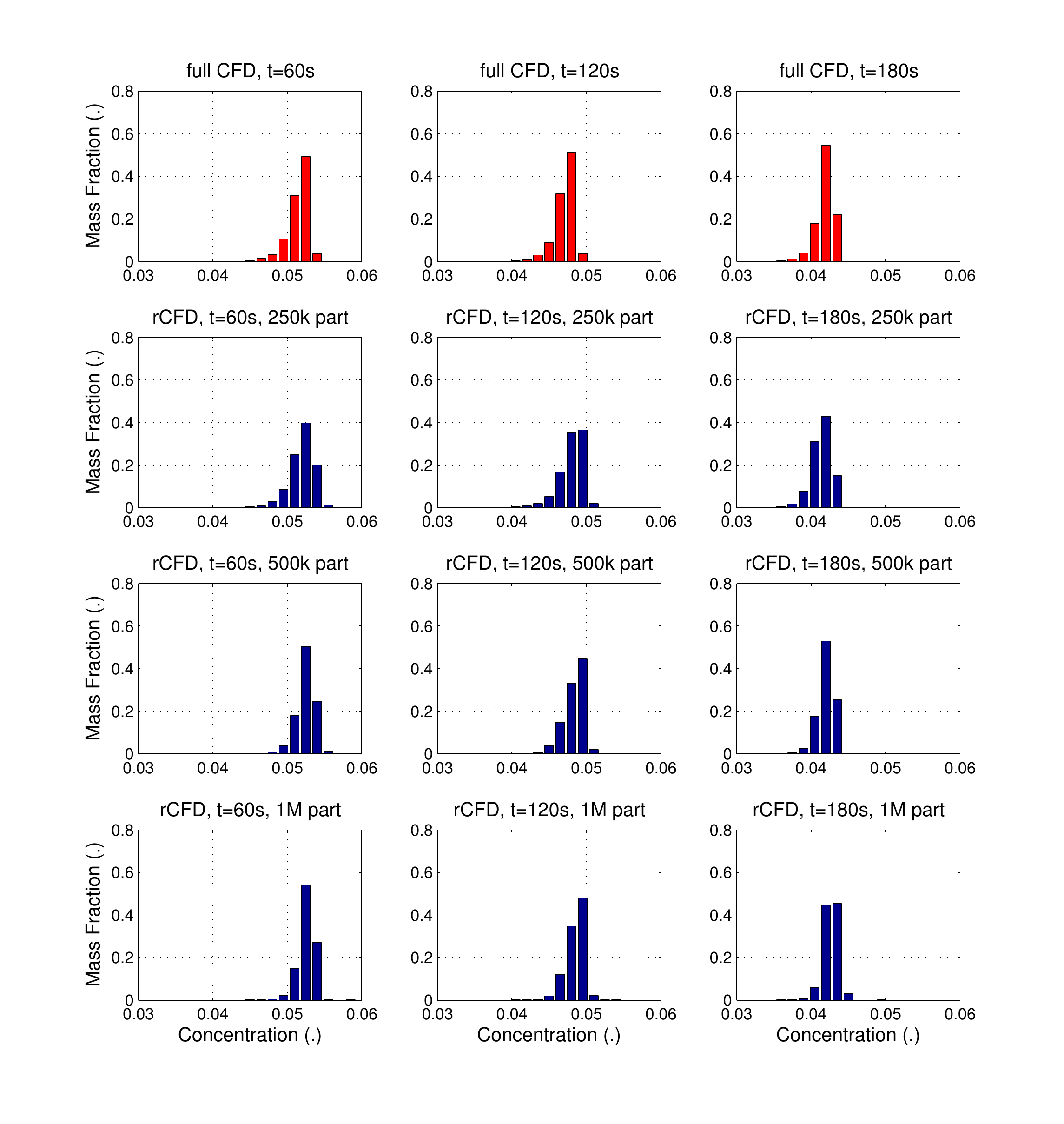}
\caption{Histograms of carbon concentration in liquid steel after one, two and three minutes of process time (columns) as a result of full (top row) and recurrence CFD at different resolutions (three bottom rows).}
\label{fig:concplot4}
\end{figure*}

Finally, in Fig.~\ref{fig:concplot4} the evolution of melt homogenization is depicted by histograms for the carbon concentration. Once again the predictions of recurrence CFD at different levels of discretization are in fairly good agreement with full CFD simulations. \myDelete{It should be noted that the concentration binning has been chosen arbitrarily. An optimized binning would lead to an improved representation of agreement, especially in case of the highest resolution of recurrence CFD.}\myHighlight{However, in this case a resolution of 500k particles seems to be superior to an increased resolution of 1M particles. At this point, we believe that these discrepancies might be associated with the ratio between the particle size and the underlying Eulerian grid spacing. In case of small particles, the mapping from large cells (with associated large reaction turnover) to small fluid particles might cause some problems. Consequently, the resolution dependence of our Eulerian/Lagrangian mapping scheme should be addressed in future.}

On top of this successful application of recurrence CFD in its Lagrangian Model B version, we emphasize that these prediction came along with a dramatic decrease in computational costs. Without any optimization in the implementation, overall simulation times were reduced by two orders of magnitude. Wall-clock times for recurrence CFD runs ranged from less than an hour at low resolution to about three hours at high resolution, in contrast to two weeks of wall-clock time required for full CFD simulations.

This application example of a steelmaking BOF converter proves that complex multiphase flows can be time-extrapolated by recurrence CFD for the purpose of addressing long-term processes. In our example we successfully considered steel decarbonisation by time-extrapolating a highly dynamic multiphase flow of a bubble-stirred melt. With the help of recurrence CFD, we could bridge the gap between the small time-scales of the flow dynamics and the large time-scale of the chemical process.
% ----------------------------------------------------------------------
\section{Conclusion and outlook}
In order to simulate flows with very distinct time scales more efficiently, we introduced a new modelling approach in this study. First, a flow's degree of recurrence is quantified and visualized using recurrence statistics and plots based on data of a conventional CFD simulation. In a second step, flow field sequences of arbitrary length are created based on information contained in the recurrence plot. Connecting similar states of the initial simulation, these chains approximate the true evolution of a sufficiently recurrent system. In a third phase, a passive process can be simulated on the flow sequences without costly solution of the underlying, fast dynamics. This way, long-term properties of such systems can be addressed.

We proposed two methods to carry out calculations on recurrence fields, a Eulerian (Model A) and a Lagrangian (Model B) one. In Model A, a passive transport equation, e.g.\ for species transport, is solved using volume fraction and velocity fields from the recurrence data base. In Model B, we trace particles (either obtained from discretization of a continuous phase or physical particles) on recurrence velocity fields and let them carry a passive species, temperature etc. For highly recurrent systems, Model A leads to almost identical results as conventional simulations with only tiny fractions of their run times. Depending on the number of particles, Model B shows a similarly strong performance as Model A, but is less suitable to resolve small-scale spatial structures. However, it can also be used in weakly recurrent situations, making it useful for more realistic scenarios such as industrial processes.

Since our aim was to introduce a new modelling approach in a basic formulation, several open questions and future challenges can be identified. In our opinion, there are at least two main weaknesses present in the current version: Firstly, the need for conventional CFD simulations to create recurrence statistics and data bases imposes bottlenecks albeit much less restrictive ones than with ordinary simulations coupling fast and slow processes. Secondly, we have to assume the slow degrees of freedom to be completely passive, i.e.\ they do not couple back to the fast ones, otherwise our recurrence fields would lose their validity over time.

While the time-consuming creation of recurrence statistics cannot be completely circumvented, the amount of required data may be reduced by optimization of the path through the recurrence statistics. The second issue might be addressed by carrying out conventional simulations corresponding to different process states to obtain several recurrence data bases for progressing system evolution.
 
Finally, we stress that given the very good performance of both Models A and B on a desktop computer and without any algorithmic optimizations, we are convinced that recurrence CFD has the capability to enable real-time simulations of industrial-scale.

\section*{Nomenclature}
\subsection*{Greek letters}
\begin{tabular}{m{1cm} l }
$\alpha$  & volume fraction \\
$\delta\alpha$ & excess of volume fraction \\
$\Delta t$ & duration of time interval \\
$\boldsymbol{\phi}$ & volumetric flux \\
$\dot{\gamma}$ & shear rate \\
$\kappa$ & mass transfer coefficient \\
$\tau_{\text{f}}$ & random walk time step \\
$\tau_{\text{p-p}}$ & (pseudo)period's duration \\
$\tau_{\text{rec}}$ & recurrence statistics' monitoring period \\
\end{tabular}
\subsection*{Latin of symbols}
   \begin{tabular}{m{1cm} l }
  $A_{\text{r}}$ & reaction area \\
  $c$ & scalar concentration \\
  $C$ & global amount of scalar quantity \\
  $D$ & diffusion coefficient \\
  $d_{\text{b}}$ & bubble diameter \\
    $d\rvec$& total displacement \\
   $d\textbf{w}$& displacement due to fluctuations  \\
   $f_{\text{p-p}}$& (pseudo)period's frequency \\
   $f_{\text{rec}}$& recurrence statistics' monitoring frequency \\
   $g$& averaging function \\
   $\textbf{j}_{\text{diff}}$& diffusive flux \\
   $l_{\text{f}}$& random walk step size \\
   $n_{\text{p}}$& number of particles per cell \\
   $\dot{N}$& molar reaction rate \\
   ${\cal N}$& recurrence norm's normalization constant \\
   $\rvec$& spatial position \\
   $R$& recurrence norm
   \end{tabular}
   
   \begin{tabular}{m{1cm} l }
  
   $R_{m,n}$& recurrence matrix \\
   $S$& scalar source term \\
  $t^{\text{(b)}}$ & begin time of interval \\
  $t^{\text{(e)}}$ & end time of interval \\
  $t^{\text{(elap)}}$ & elapsed time after interval \\
  $\uvec$ & velocity \\
  $V_c$ & volume of cell $c$ \\
  $V_{\text{p}}$ & particle volume
\end{tabular}
\subsection*{Subscripts/Superscripts}
\begin{tabular}{m{1cm} l }
     $\alpha$& related to volume fraction \\
     eq& in equilibrium \\
     $i$& particle or interval $i$ \\
     $(i)$& phase $i$ \\
     m& metal-quantity \\
     $\phi$& related to flux \\
     rec& related to recurrence statistics \\
     s& slag-quantity
\end{tabular}
 \subsection*{List of abbreviations}
\begin{tabular}{m{1cm} l }
   BOF& basic oxygen furnace \\
   Co& Courant number \\
   CFD& computational fluid dynamics \\
   DEM& discrete element method \\
   LES& large eddy simulation \\
   POD& proper orthogonal decomposition \\
   ROM& reduced order model \\
   RQA& recurrence quantification analysis \\
   VoF& volume of fluid
\end{tabular}

\section*{Acknowledgement}
We acknowledge support from K1-MET GmbH metallurgical competence center and code contributions from Gerhard Holzinger.

%% References
\section*{References}


%\bibliographystyle{elsarticle-num}
\bibliographystyle{model2-names.bst}\biboptions{authoryear}
\bibliography{rCFDpaper}

 
\end{document}